\documentclass[%
reprint,
 amsmath, amssymb,
 aps, floatfix
]{revtex4-2}

\usepackage{graphicx}
\usepackage{dcolumn}
\usepackage{bm}
\usepackage[colorlinks]{hyperref}
\usepackage{braket}
\usepackage{physics}
\usepackage{amsmath}
\usepackage{array}
\usepackage{float}

\begin{document}
\preprint{APS/123-QED}
\title{Harnessing quantumness of states using discrete Wigner functions under (non)-Markovian quantum channels}
\author{Jai Lalita}
\email{jai.1@iitj.ac.in}
\author{K. G. Paulson}%
\email{paulsonkgeorg@gmail.com}
\author{Subhashish Banerjee}
\email{subhashish@iitj.ac.in }
\affiliation{Indian Institute of Technology, Jodhpur-342030, India}%

\date{\today}

\begin{abstract}
The negativity of the discrete Wigner functions (DWFs)  is a measure of non-classicality and is often used to quantify the degree of quantum coherence in a system. The study of Wigner negativity and its evolution under different quantum channels can provide insight into the stability and robustness of quantum states under their interaction with the environment, which is essential for developing practical quantum computing systems. We investigate the variation of DWF negativity  of qubit, qutrit, and two-qubit systems under the action of (non)-Markovian  random telegraph noise (RTN) and amplitude damping (AD) quantum channels. We construct different negative quantum states which can be used as a resource for quantum computation and quantum teleportation. The success of quantum computation and teleportation is estimated for these states under (non)-Markovian evolutions.
\end{abstract}
\maketitle
\section{\label{sec:level1}Introduction\protect}
The notion of phase space is essential in investigating classical systems' dynamics. The uncertainty principle, however, limits its straightforward application to the quantum scenario. Nevertheless, it is still conceivable to create quasi-probability distributions (QDs) for quantum mechanical systems analogous to their classical counterparts. Wigner created the first QD, now known as the Wigner function (WF) \cite{wigner1932quantum, hillery1984distribution}. It is not only real-valued and normalized but also gives the correct value of the probability density for the quadrature $a\textbf{Q} + b\textbf{P}$ (here $\textbf{Q}$ and $\textbf{P}$ are canonical positions and momentum operators) when integrated along the phase-space line $aq + bp$. However, unlike probability densities, the Wigner function can assume negative values for some quantum states; thus making it a quasi-probability distribution function. Classical light states, like coherent states, have positive Wigner functions \cite{hudson1974wigner}, whereas quantum light states, like photon-added/, subtracted coherent and entangled states, do not \cite{zavatta2004quantum, meena2022characterization, malpani2020impact}. Negative Wigner function value is a witness of the quantum nature of the state \cite{kenfack2004negativity}. Through homodyne measurements, Wigner functions can be experimentally recreated, and the visual representation of the recreated state effectively highlights quantum interference processes \cite{leonhardt1997measuring}.

 The original Wigner function only applies to the continuous situation, whereas density operators can also express discrete degrees of freedom like spin. Therefore, given the significance of Wigner functions for continuous variable (CV) systems \cite{agarwal1981relation,agarwal1998state,thapliyal2015quasiprobability,thapliyal2016tomograms}, much emphasis has been paid to creating their finite-dimensional analogues as we generally deal
with finite-dimensional Hilbert space systems in quantum information and processing. For example, for a system of $n$ qubits, the dimension of the Hilbert space of states is $d = 2^n$. For such systems, various discrete analogues of the Wigner function have been proposed \cite{cohen1986joint,wootters1987wigner,galetti1988extended,leonhardt1996discrete,wootters2004picturing,gibbons2004discrete,chaturvedi2005wigner}. The Wigner function formulation applied to arbitrary spin systems of prime dimensions was developed in \cite{wootters1987wigner}. It is defined on an explicitly geometrical phase space over a finite mathematical field. The integers $0,..., d-1$, with addition and multiplication mod $d$, make up the finite mathematical field (where $d$ is the dimension of the system's Hilbert space). Later, this formulation has been redressed to the power of prime dimension \cite{wootters2004picturing, gibbons2004discrete}. A tomographical scheme was proposed to infer the quantum states of finite—dimensional systems from experiments by developing a new discrete Wigner formalism \cite{leonhardt1996discrete}. An algebraic approach was provided to find the Wigner distributions for finite odd-dimensional quantum systems \cite{chaturvedi2005wigner}. Withal, discrete Wigner functions (DWFs) have been used to investigate a variety of exciting problems connected with quantum computation, such as magic state distillation \cite{howard2014contextuality, veitch2014resource, schmid2022uniqueness}, separability \cite{pittenger2005wigner}, quantum state tomography \cite{wootters2004picturing, paz2004quantum}, teleportation \cite{koniorczyk2001wigner, paz2002discrete}, decoherence \cite{lopez2003phase}, error correction \cite{paz2005qubits}. 

Here we focus on the class of DWFs defined in \cite{wootters2004picturing, gibbons2004discrete} for power-of-prime dimensions. This class defines DWFs by associating lines in discrete phase space to projectors belonging to a fixed set of mutually unbiased bases (MUBs) detailed in Sec. \ref{prelim}. The advantage of this formulation is that DWFs transparently correspond to the expectation values of the phase space point operators. These phase space point operators are built using quantum states connected to phase space lines \cite{wootters1987wigner,gibbons2004discrete}. Tomographic reconstruction of this class of DWFs of entangled bipartite photonic states using MUBs is presented in \cite{srinivasan2017investigations}. The method of extremizing DWFs was contemplated by finding states corresponding to normalized eigenvectors of the minimum and maximum eigenvalues of phase space point operator \cite{casaccino2008extrema}. Thereafter this idea was extended for all odd prime dimensions to find the maximally negative quantum states \cite{van2011noise}, $\textit{i.e.}$, states corresponding to a minimal eigenvalue of phase space point operator's normalized eigenvector. It was shown that these states are maximally robust toward depolarizing noise. 

In this work, we focus on the negative quantum states of a qubit, qutrit, and two-qubit systems to examine how their DWFs and discrete Wigner negativity $|N_{G}(\pmb{\rho})|$ change under the impact of a variety of noisy channels, both unital and non-unital, in the (non)-Markovian regimes.   
Noise emerges as an artefact of the system's interaction with its ambient environment.
 The theory of open quantum systems (OQS) \cite{nielsen2002quantum,breuer2002theory,banerjee2018open,weiss2012quantum,czerwinski2022dynamics} offers a framework for investigating how an environment affects a quantum system. Ideas of open quantum systems have wide applicability \cite{caldeira1983quantum,grabert1988quantum,hu1994quantum,banerjee2000quantum,banerjee2003general,srikanth2008squeezed,plenio2008dephasing,omkar2014unruh,iles2014environmental,banerjee2017characterization,banerjee2017characterization,paulson2016hierarchy,naikoo2018study,tanimura2020numerically,teklu2022noisy,czerwinski2022efficiency}. In many cases, the dynamics of an OQS can be described using a Markovian approximation where a clean separation between the system and environment time scales exist. When this is not so, we enter the non-Markovian regime \cite{de2017dynamics,rivas2014quantum,li2018concepts,vacchini2012classical,breuer2016colloquium,daffer2004depolarizing,kumar2018non,naikoo2020coherence,utagi2020ping,naikoo2019facets,shrikant2018non,utagi2020temporal,thomas2018thermodynamics,paulson2021hierarchy}.

With the motivation to understand the impact of noise on the DWFs under the action of (non)-Markovian, unital (illustrated by the random telegraph noise) as well as non-unital (depicted by the amplitude damping noise), we calculate the DWFs for the qubit, qutrit, and two-qubit systems. In particular, we study the variation of DWFs corresponding to the maximally negative quantum state, $\textit{i.e.}$, the first negative quantum ($NS_{1}$) state of the qubit, qutrit, and two-qubit systems under the same (non)-Markovian noisy channels. Negative quantum states are discussed in greater detail in Sec. \ref{negative state}. A concept connected with the states having negative discrete Wigner functions is the mana \cite{veitch2014resource}, which has been used in the literature to compute magic associated with non-stabilizer states. Magic states are found to be ideal resources for quantum computational speedup and fault-tolerant quantum computation. Here, we compute the mana of the qutrit's first and second negative quantum states \cite{jain2020qutritmagic}. We also study the variation of mana under the aforementioned (non)-Markovian channels. We also examine discrete Wigner negativity for the power of prime dimension systems (for $d = 2, 3, 2^2$) under the same (non)-Markovian channels. Quantum coherence is a fundamental prerequisite for all quantum correlations, including entanglement, and it is a crucial physical resource in quantum computation and information processing \cite{baumgratz2014quantifying, xi2015quantum, streltsov2017colloquium, hu2018quantum, zhao20191,paulson2022quantum}.
 Also, entanglement is a premium quantum correlation and has many operational uses. We will study the dynamics 
of quantum coherence and entanglement, \textit{i.e.}, concurrence \cite{wootters1998entanglement}  utilizing two-qubit's first, second, and third negative quantum state using DWFs and compare them with the corresponding dynamical evolution of Bell states. Average fidelity is commonly used to gauge a channel's performance \cite{horodecki1996teleportation}. The notion of fidelity is a qualitative metric for differentiating between two quantum states \cite{ghosal2021characterizing, naikoo2020coherence}. We will use DWFs to compare the average fidelity of the two-qubit system's first, second, and third negative quantum states with the Bell states when subjected to the above noisy channels.  

 The paper is organized as follows. In Sec. \ref{prelim}, we discuss the essential ingredients of the DWFs. Sections \ref{DWF-noise} and \ref{neg} study the behaviour of DWFs of quantum systems (particularly qubit, qutrit, and two-qubit) and variation of discrete Wigner negativity $|N_G(\pmb{\rho})|$ and mana, under (non)-Markovian unital (random telegraph noise) and non-unital (amplitude damping) channels. In Secs. \ref{coh, con} and \ref{fid}, we study the variation of quantum coherence, concurrence, and fidelity for the first, second, and third negative quantum state and Bell state under the same (non)-Markovian channels, followed by the conclusion in Sec. \ref{conclusion}. 
\section{\label{prelim}Preliminaries} 
This section briefly reviews the concept of discrete phase space and mutually unbiased bases. Further, we present the class of DWFs for the power of prime dimensions proposed in \cite{wootters2004picturing, gibbons2004discrete}, followed by a discussion on negative quantum states and the discrete counterpart of non-classical volume (mana). 
\subsection{Discrete phase space and mutually unbiased bases} $\label{stria}$
A $d \times d$ real array is the discrete equivalent of phase space in a $d$-dimensional Hilbert space. Suppose we are describing a quantum state whose dimension $d$ in Hilbert space is a power of a prime ($d = p^n$). In these circumstances, label the position and momentum coordinates of the $d \times d$ grid with finite Galois field GF($p^n$) elements \cite{lidl1994introduction}. This is because if we do so, we can endow the phase-space grid with the same geometric properties as the ordinary plane. For instance, we can define the finite $d \times d$ grid linear equations of the form $aq + bp = c$ and lines as its solutions (where all the components and actions in this equation are contained in GF ($p^n$) ). In contrast to the continuous phase space, this discrete arrangement has no geometrical lines of points. Instead, a line is a collection of $d$ distinct phase space points, depicted by the same colored dots in Figs. (\ref{striation1}), (\ref{striation2}), and (\ref{striation3}). Our discrete phase space may then be divided into various parallel line collections. Each of these collections is referred to as a striation \cite{wootters2004picturing}. A method for creating $d + 1$ striations of a $d \times d$ phase-space array as described in \cite{gibbons2004discrete}, produces striations with the three following properties:\\
$(i)$ For given pair of points in the discrete phase space, there is exactly one line containing both points.\\
$(ii)$ Two non-parallel lines intersect exactly at one point, $\textit{i.e.}$, they share only one common point.\\
$(iii)$ For any phase-space point $\alpha(q, p)$ which is not contained in the line $\lambda$, there is exactly one line parallel to $\lambda$ containing the phase-point $\alpha(q, p)$.\\
In Sec. \ref{DWF}, we will see that these striations are crucial for formulating the discrete Wigner function $W$.

A unique set of ($d + 1$) bases is used for a $d$-dimensional Hilbert space to define DWFs. If the dimension of the space of states is a power of a prime integer, then it is known that there exists a complete set of $(d + 1)$ MUBs. Let's consider two distinct orthonormal bases, $B_{1}$ and $B_{2}$, such that
\begin{equation}
B_{1} = \{\ket{\beta_{1,1}},\ket{\beta_{1,2}},....,\ket{\beta_{1,d}}\}, |\bra{\beta_{1,i}\ket{\beta_{1,j}}}^2 = \delta_{i,j}, 
\end{equation}
\begin{equation}
B_{2} = \{\ket{\beta_{2,1}},\ket{\beta_{2,2}},....,\ket{\beta_{2,d}}\}, |\bra{\beta_{2,i}\ket{\beta_{2,j}}}^2 = \delta_{i,j}, 
\end{equation}
These are mutually unbiased if,
\begin{equation}
|\bra{\beta_{i,j}\ket{\beta_{i',j'}}}^2 = \frac{1}{d}(1 - \delta_{i,i'}) + \delta_{i,i'}\delta_{j,j'}.
\end{equation}
Note that this is exactly the number of striations one can find with properties (i)–(iii) above in Sec. \ref {stria}. Numerous MUB constructions have been suggested in the literature \cite{wootters1989optimal,
lawrence2002mutually, bandyopadhyay2002new, pittenger2004mutually, lawrence2002mutually, bandyopadhyay2002new, pittenger2004mutually} for such situations.
\subsection{\label{DWF} Discrete Wigner functions (DWFs)} 
Now, we have a collection of $(d + 1)$ mutually unbiased bases $(B_1, B_2,..., B_{d+1})$ and a set of $(d + 1)$ striations $(S_1, S_2,..., S_{d+1})$ of the $(d \times d)$ phase space having $d$ parallel lines. We must select two one-to-one mappings to build a DWF, $\textit{i.e.}$, each basis set $B_i$ is connected to one striation $S_i$, and each basis vector $\ket{\beta_{i, j}}$ is connected to a line $\lambda_{i,j}$ (the $j^{th}$ line of the $i^{th}$ striation).  Each line $\lambda_{i, j}$ of $i^{th}$ striation is connected to a projector $\textbf{P}_{i,j} = \ketbra{\beta_{i,j}}{\beta_{i,j}}$ onto a basis state of striation to define the quantum net. This may be accomplished in various ways, each producing a different quantum net and, thus, a different definition of the discrete Wigner function $W$, as detailed in \cite{gibbons2004discrete}. Given that these linkages exist, the DWFs are uniquely defined as 
\begin{equation}
p_{i,j} \equiv \Tr[\ketbra{\beta_{i,j}}{\beta_{i,j}}\pmb{\rho}] = \sum_{\alpha \in \lambda_{i,j}} W_{\alpha}, 
\end{equation}
In other words, we want the probability of projecting onto the basis vector corresponding to each line to be equal to the sum of the Wigner function elements corresponding to that line. Then it can be demonstrated that the resultant Wigner function at any phase-space point $\alpha(q, p)$ is \cite{gibbons2004discrete},
\begin{equation}
    \begin{aligned}
      W_{\alpha} = \frac{1}{d} \Tr[ \textbf{A}_{\alpha} \pmb{\rho} ],
    \end{aligned}\label{DWFformula}
\end{equation}
where
\begin{equation}
    \begin{aligned}
      \textbf{A}_{\alpha} = \sum_{\alpha \in \lambda_{i,j}} \textbf{P}_{i,j} - \textbf{I}.
    \end{aligned}\label{A_formula}
\end{equation}
The operators $\textbf{A}_{\alpha}$, known as phase-space point operators, are Hermitian operators. Therefore, all their eigenvalues are real. They also form a complete basis for the space of operators, which are orthogonal in the Schmidt inner product (i.e.,  $Tr[\textbf{A}_{\alpha}\textbf{A}_{\beta}] = \delta_{\alpha,\beta}/d$ ) and $Tr(\textbf{A}_{\alpha}) = 1$. The sum of phase space point operators $\textbf{A}_{\alpha}$ along any line $\lambda$ is equal to the projectors associated with it, $\textit{i.e.}$, $\textbf{P}(\lambda) = \sum_{\lambda \ni \alpha} \textbf{A}_{\alpha}$. Any density operator can also be written as, 
\begin{equation}
    \begin{aligned}
      \pmb{\rho} = \sum_{\alpha} W_{\alpha}\textbf{A}_{\alpha},
    \end{aligned}\label{rho-decomposition-in-A}
\end{equation}
where $W_{\alpha}$ are the expansion coefficients. It can be shown that the discrete Wigner function $W$ shares many characteristics with the continuous Wigner function $W(q, p)$ \cite{gibbons2004discrete} such as it is real (but can also be negative), normalized, and gets its values from measurements onto MUB using Eq. (\ref{DWFformula}). In this case, the MUB projectors take on the role that the quadratures $a\textbf{Q} + b\textbf{P}$ play in $W(q, p)$, providing a highly symmetric collection of observables whose measurement results fully define the state (quantum tomography). In Refs. \cite{gibbons2004discrete, paz2005qubits, pittenger2004mutually, galvao2005discrete, cormick2006classicality}, more features of $W$ are examined.

\subsection{\label{negative state} Negative quantum states}
If a state $\pmb{\rho}$ is outside the convex hull of stabilizer states, $\textit{i.e.}$, simultaneous eigenstates of generalized Pauli operators \cite{gottesman1997stabilizer},  $\Tr[ \textbf{A}_{\alpha} \pmb{\rho} ]$ is less than zero for at least one of the phase-space operators $\textit{A}_{\alpha}$. The discrete Wigner negativity $|N_{G}(\pmb{\rho})|$ of a state $\pmb{\rho}$, using DWFs, is defined in the following way \cite{van2011noise}, 
 \begin{equation}
 |N_{G}(\pmb{\rho})| = 
 \begin{cases}
    \left| min_{\alpha \in Z^{D+1}_D} { (\Tr[\textbf{A}_{\alpha}\pmb{\rho}])}\right|, &   \Tr[\textbf{A}_{\alpha}\pmb{\rho}] < 0, \\
      0, &   \Tr[\textbf{A}_{\alpha}\pmb{\rho}] \geq 0.  
\end{cases}\label{negativity}
\end{equation}
Furthermore, an expression for the robustness of $D$-prime dimensional states toward depolarizing noise having error probability $a$, using the discrete Wigner negativity of states $|N_G(\pmb{\rho})|$ was developed in \cite{van2011noise},
\begin{equation}
    \begin{aligned}
      a^{*}(\pmb{\rho}) = 1 - \frac{1}{D^2 |N_{G}(\pmb{\rho})| + 1},  
      \end{aligned}\label{robust}
\end{equation}
where,
$a^{*}(\pmb{\rho})$ is $min(a)$ such that
\begin{equation}
    \begin{aligned}
      (1 - a)\pmb{\rho} + a\frac{I}{D} = \sum_{i} c_{i}\ketbra{S_i}. 
      \end{aligned}
\end{equation}
Here $0 \leq c_i \leq 1$,  $ \sum_{i}c_{i} = 1$ and $\ket{S_i}$ are D-dimensional stabilizer states.
Using Eq. (\ref{robust}), it was shown that for all $D$-prime dimensions, maximally negative quantum states are maximally robust to depolarizing noise.\\ 
We will take this up for the negative quantum states of the power of prime dimensional systems ($d = p^n$), especially qubit, qutrit, and two-qubit systems and their corresponding DWFs to study the variation of DWFs for a wide range of noisy channels, both unital and non-unital, in the (non)-Markovian regimes. The variation of discrete Wigner negativity, $\textit{i.e.}$, $|N_{G}(\pmb{\rho})|$ for these systems is also studied under the same noisy channels. Additionally, we will compare the concurrence and teleportation fidelity of the two-qubit negative quantum states with that of the Bell states. These maximally negative quantum states are calculated by minimizing $Tr(\pmb{\rho}\textbf{A}_{\alpha})$, as elaborated in \cite{casaccino2008extrema,van2011noise}, by finding the minimum eigenvalue of $\textbf{A}_{\alpha}$, and using the corresponding normalized eigenvector for $\pmb{\rho}$. In our discussion, we denote the maximal negative quantum state as the first negative quantum ($NS_1$) state. The second negative quantum ($NS_2$) state and third negative quantum ($NS_3$) state correspond to normalized eigenvectors of the second and third negative eigenvalues of $\textbf{A}_{\alpha}$, respectively, and so on.
\subsection{\label{magic state} Discrete counterpart of non-classical volume}
The mana $\pmb{M}(\pmb{\rho})$ of a state $\pmb{\rho}$ provides information about its applicability in magic state distillation protocols \cite{veitch2014resource}. It is defined as
\begin{equation}
    \begin{aligned}
     \pmb{M}(\pmb{\rho}) \equiv \log \left[ \sum_{\alpha} |W_{\alpha}| \right] = \log(2 Sn(\pmb{\rho}) + 1),  
      \end{aligned}\label{mana}
\end{equation}
where, $Sn(\pmb{\rho})$ is sum negativity given by,
\begin{equation}
    \begin{aligned}
     Sn(\pmb{\rho}) \equiv \sum_{\alpha:W_{\alpha}<0}|W_{\alpha}| \equiv \frac{1}{2} \left(\sum_{\alpha}| W_{\alpha}| - 1\right).  
      \end{aligned}
\end{equation}

A physical interpretation of mana is provided by $Sn(\pmb{\rho})$. It is the absolute value of sum negative entries in the DWFs of a state $\pmb{\rho}$. These negative entries are a hindrance to classical computation and hence motivate quantum computation. Further, it is the discrete counterpart of the nonclassical volume \cite{kenfack2004negativity,thapliyal2015quasiprobability,teklu2015nonlinearity}, defined using the Wigner function in the CV regime.
\section{\label{DWF-noise}DWFs of quantum systems under noisy channels}  
In this section, we calculate the DWFs for single-qubit, single-qutrit, and two-qubit systems, using the formalism given in Sec. \ref{DWF}. We then identify the DWFs for the first negative single-qubit, single-qutrit, and two-qubit quantum states. Further, we examine their fluctuations under various Markovian and non-Markovian channels.
\subsection{Single qubit} 
\begin{figure}[!htpb]
    \centering
    \includegraphics[width = 0.4\textwidth, height = 75mm]{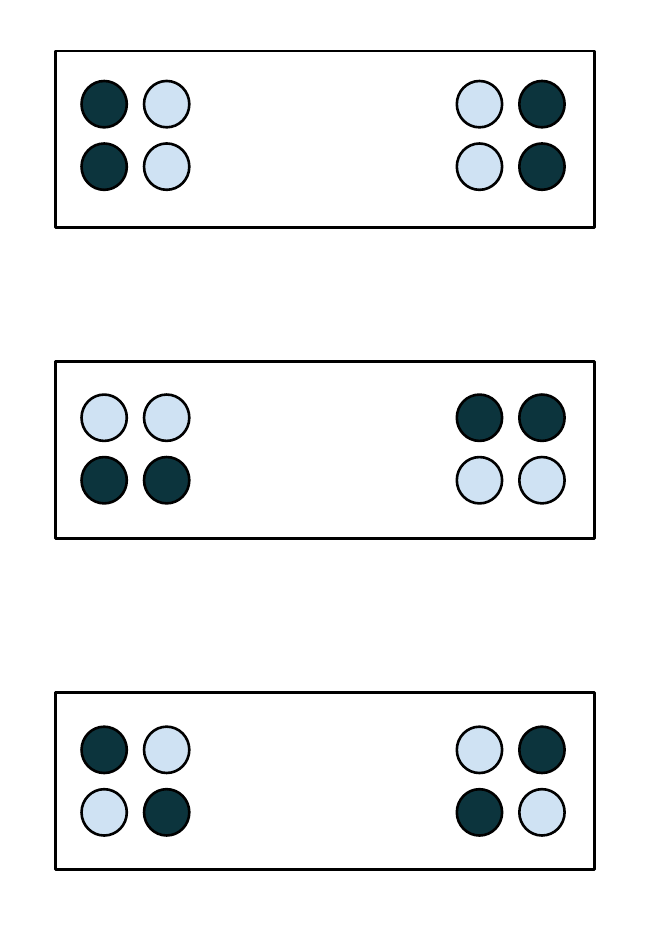}
    \caption{Lines and striations of the $2 \times 2$ phase space.}
    \label{striation1}
\end{figure}

$\label{single-qubit}$
The discrete phase space for a single-qubit system is defined on a $2 \times 2$ real array. The points in this discrete phase space are labeled by elements of the Galois field $\mathcal{F}_2 = \{0, 1\}$. The eigenstates of Pauli operators, $\sigma_x$, $\sigma_y$, and $\sigma_z$ can conveniently be chosen as MUBs for single-qubit systems \cite{galvao2005discrete}. This phase space has three striations, each having the necessary properties $(i), (ii)$, and $(iii)$ listed in Sec. \ref {stria} and displayed in Fig. (\ref{striation1}). A collection of three mutually unbiased bases must now be defined for one-to-one mapping with striations, as seen from TABLE \ref{table1}.

Using the Bloch vector representation for a single-qubit system, {\it i.e.}, $\pmb\rho = \frac{1}{2}( \textbf{I}_2 + \textbf{a} . \pmb{\sigma})$, (here $\textbf{a} \in \textbf{R}^3$ and $\pmb{\sigma}$'s are Pauli spin matrices) in Eq. (\ref{DWFformula}), we find the expressions for single-qubit DWFs for a given association of MUB's as  
\begin{eqnarray}
\nonumber
W_{1, 1} = \frac{1}{4} (1 - a_2 + a_3),
 W_{1, 2} = \frac{1}{4} (1 + a_2 - a_3),\\ 
 W_{2, 1} = \frac{1}{4} (1 + a_2 + a_3),
W_{2, 2} = \frac{1}{4} (1 - a_2 - a_3), 
\end{eqnarray}
where $a_1$, $a_2$, and $a_3$ are the components of the single-qubit Bloch vector $\textbf{a}$.
\begin{table}
\begin{tabular}{ | m{2cm}| m{5cm} | }
  \hline
  \textbf{Striation} & \textbf{MUBs associated with striation}\\ 
  \hline
  1 & 
  $\begin{pmatrix}
  0\\
  1
\end{pmatrix}$,
$\begin{pmatrix}
  1\\
  0
\end{pmatrix}$\\
\hline
  2 &
  $\frac{1}{\sqrt{2}}\begin{pmatrix}
  1\\
  1
\end{pmatrix}$,
$\frac{1}{\sqrt{2}}\begin{pmatrix}
  1\\
  -1
\end{pmatrix}$\\
\hline
  3 & 
  $\frac{1}{\sqrt{2}}\begin{pmatrix}
  1\\ 
  \iota
\end{pmatrix}$,       
$\frac{1}{\sqrt{2}}\begin{pmatrix}
  1\\ 
  -\iota
\end{pmatrix}$\\
\hline
\end{tabular}
\caption{\label{table1} The MUBs associated with lines of the $2 \times 2$ discrete phase space of single-qubit systems.}
\end{table}
\subsubsection{\label{RTNqubit} Random Telegraph Noise}
When a system is exposed to a bi-fluctuating classical noise that generates random telegraph noise (RTN) with pure dephasing \cite{daffer2004depolarizing, kumar2018non}, this channel characterizes the system's dynamics. We try to understand how the single-qubit's $NS_{1}$ state DWFs evolve in the presence of (non)-Markovian random telegraph noise (RTN). The dynamical map for a single-qubit system under the action of (non)-Markovian RTN channel is:
\begin{equation}
    \begin{aligned}
      \epsilon^{RTN}(\pmb{\rho}) = \mathbf{R}_0\pmb{\rho}\mathbf{R}^{\dag}_0 + \mathbf{R}_1\pmb{\rho}\mathbf{R}^{\dag}_1,   
      \end{aligned}\label{RTNfinalrho}
\end{equation}
where the two Kraus operators are given as
\begin{eqnarray}
      \mathbf{R}_0 = \sqrt{\frac{1 + \Lambda(t)}{2}}\textbf{I}_{2},
      \mathbf{R}_1 = \sqrt{\frac{1 - \Lambda(t)}{2}}\pmb{\sigma_z}.
\end{eqnarray}
Here, $\Lambda(t)$ is the memory kernel
\begin{equation}
      \Lambda(t) = e^{-\gamma t}\left[ \cos\left(\zeta \;\gamma t\right) + \frac{\sin\left(\zeta \;\gamma t\right)}{\zeta}\right].
\end{equation}
Here $b$, $\gamma$ quantify the RTN's system–environment coupling strength and fluctuation rate, and $\textbf{I}$ and $\pmb{\sigma}_z$ are the identity and Pauli spin matrices, respectively, and $\zeta = \sqrt{\left(\frac{2b}{\gamma}\right)^2 - 1}$. The dynamics is Markovian if {$(4 b \tau)^2 < 1$ and non-Markovian if $(4 b \tau)^2 > 1$ and $\tau = \frac{1}{2\gamma}$ as discussed in \cite{naikoo2020coherence}.  
We employ the Bloch vector representation of single-qubit systems given in Sec. \ref{single-qubit} with Eq. (\ref{RTNfinalrho}) and Eq. (\ref{DWFformula}) for a particular association of MUBs, to determine the DWFs of a single qubit under the action of (non)-Markovian RTN channel.
\begin{eqnarray}
     W_{1, 1} &=& \frac{1}{4}\left(1 + a_3 - (a_1 + a_2)e^{-\gamma t}\cos\left({\zeta\gamma t}\right)\right. \nonumber\\
     &-& \left.\frac{(a_1 + a_2)e^{-\gamma t}\sin\left({\zeta\gamma t}\right)}{\zeta}\right),\nonumber\\
      W_{1, 2} &=& \frac{1}{4}\left(1 - a_3 + (-a_1 + a_2)e^{-\gamma t}\cos\left({\zeta\gamma t}\right) \right. \nonumber
     \\&+& \left.\frac{(-a_1 + a_2)e^{-\gamma t}sin\left({\zeta\gamma t}\right)}{\zeta}\right),\nonumber\\
      W_{2, 1} &=& \frac{1}{4}\left(1 + a_3 + (a_1 + a_2)e^{-\gamma t}\cos\left({\zeta\gamma t}\right) \right. \nonumber\\
      &+& \left.\frac{(a_1 + a_2)e^{-\gamma t}\sin\left({\zeta\gamma t}\right)}{\zeta}\right),\nonumber\\
     W_{2, 2} &=& \frac{1}{4}\left(1 - a_3 + (a_1 - a_2)e^{-\gamma t}\cos\left({\zeta\gamma t}\right) \right. \nonumber
     \\&+& \left.\frac{(a_1 - a_2)e^{-\gamma t}\sin\left({\zeta\gamma t}\right)}{\zeta}\right),
\end{eqnarray}
where $a_1$, $a_2$, and $a_3$ are the componenets of the single-qubit Bloch vector $\textbf{a}$.
Figs. (\ref{qubitNMRTN}), (\ref{qubitMRTN}), depicts the variation of the $NS_{1}$ state (for, $a_1$ = 0.50, $a_2$ = 0.56, and $a_3$ = -0.66) DWFs, of the single-qubit systems in the presence of non-Markovian and Markovian RTN noise, respectively. The characteristic oscillatory features in the non-Markovian regime can be clearly seen. 
\begin{figure}[!htpb]
    \centering
    \includegraphics[height=65mm,width=1\columnwidth]{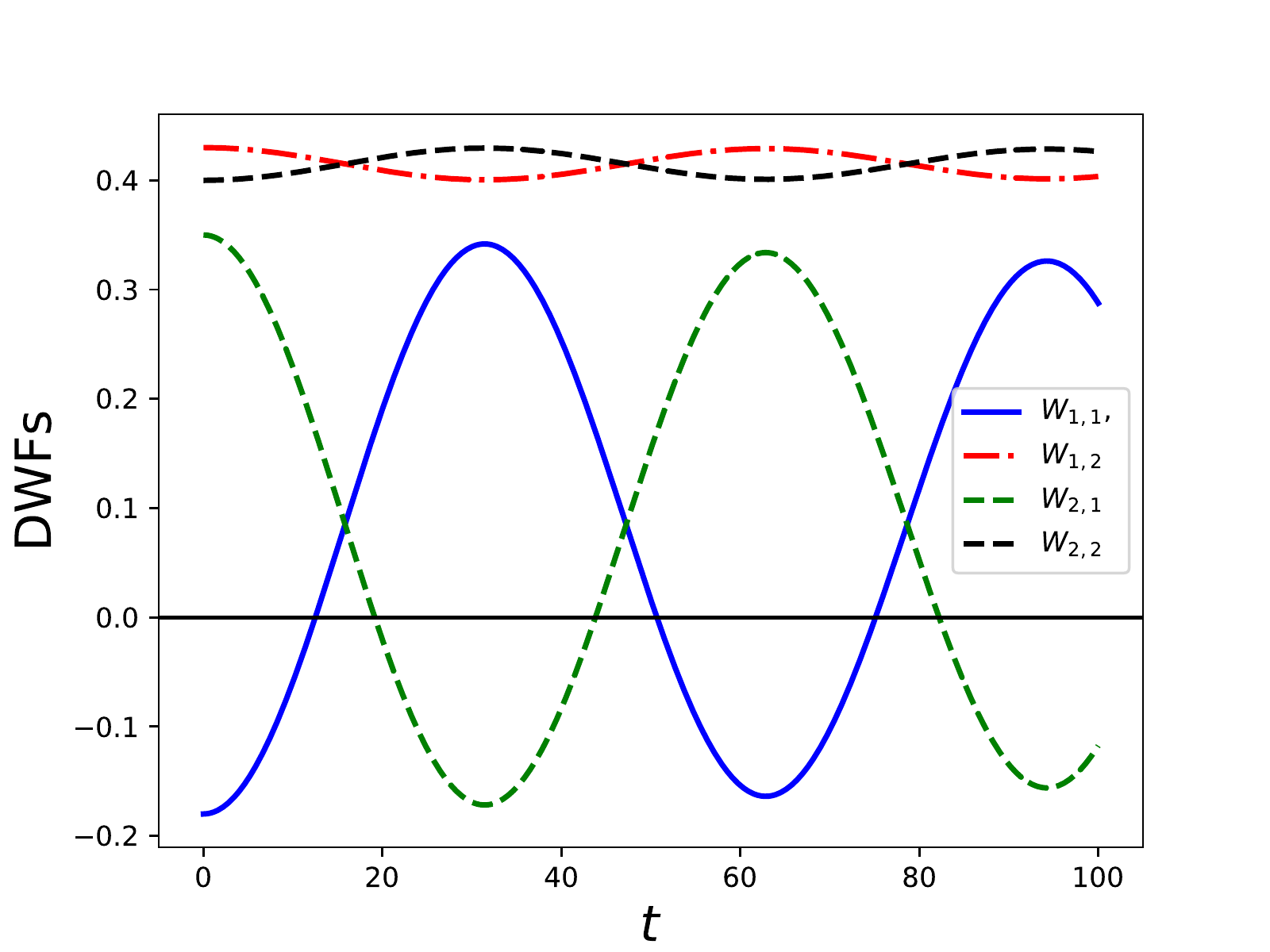}
    \caption{Variation of DWFs corresponding to the qubit's $NS_1$ state (when $a_1 = 0.50$, $a_2 = 0.56$ and $a_3 = -0.66$), under non-Markovian RTN (for $\gamma = 0.001$, $b = 0.05$) with time.}
    \label{qubitNMRTN}
\end{figure}
\begin{figure}[!htpb]
    \centering
    \includegraphics[height=65mm,width=1\columnwidth]{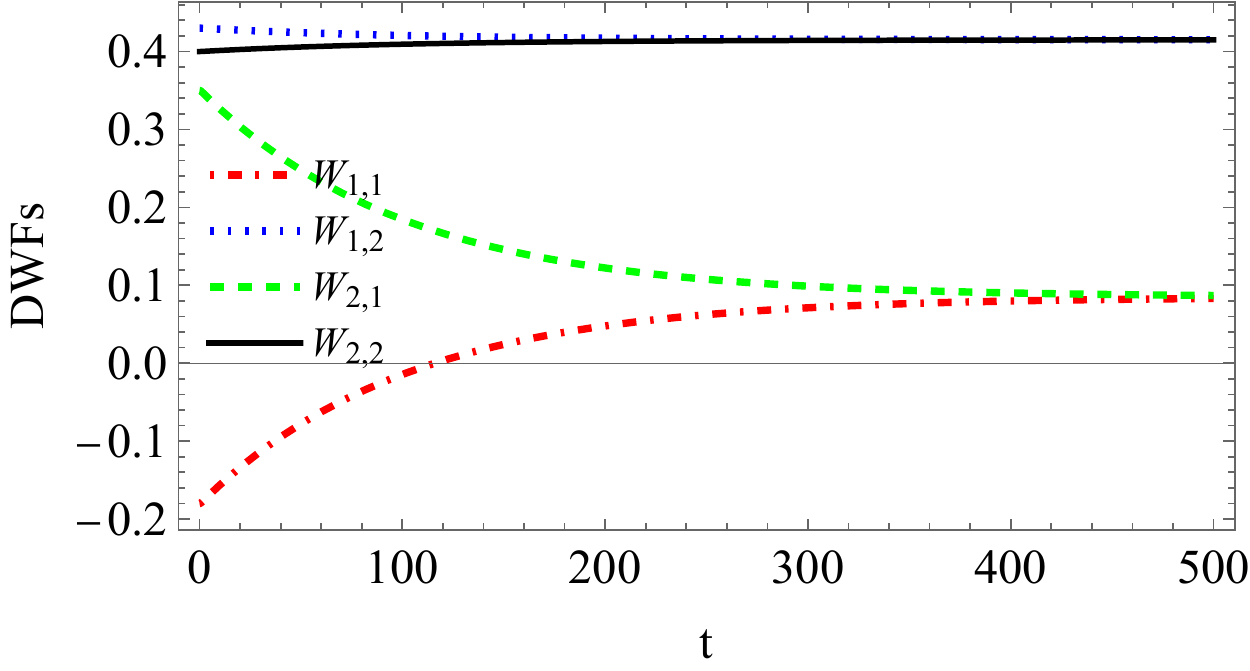}
    \caption{Variation of DWFs corresponding to the qubit's $NS_1$ state (when $a_1 = 0.50$, $a_2 = 0.56$ and $a_3 = -0.66$), under Markovian RTN (for $\gamma = 1$ and, $b = 0.07$) with time.}
    \label{qubitMRTN}
\end{figure}
\subsubsection{\label{qubitNMAD} Amplitude Damping Noise}
Here, we study how a single-qubit's $NS_1$ state DWF evolves under the influence of a  (non)-Markovian amplitude damping (AD) noise. A number of phenomena have been addressed by the application of amplitude-damping noise. This includes, among others, attenuation, energy dissipation, spontaneous photon emission, and idle errors in quantum computing in two-level systems \cite{nielsen2002quantum}. A $d$-dimensional generalization of this was introduced in \cite{dutta2016entanglement}. The following Kraus operators define the (non)-Markovian amplitude damping (AD) channel for a single-qubit system \cite{ghosal2021characterizing, utagi2020ping, dutta2021quantum}
\begin{equation}
    \mathbf{K_0} = \begin{pmatrix}
     1 & 0\\
     0 & \sqrt{1 - \lambda(t)}
    \end{pmatrix},
    \mathbf{K_1} = \begin{pmatrix}
    0 & \sqrt{\lambda(t)}\\
    0 & 0
\end{pmatrix},
\end{equation}
where, $\lambda(t) = 1 - e^{-gt}\left(\frac{g}{l} \sinh{\frac{lt}{2}} + \cosh{\frac{lt}{2}}\right)^2$, and $l = \sqrt{g^2 - 2\gamma g}$.
The system exhibits Markovian and non-Markovian evolution of a state if $2\gamma << g$ and $2\gamma >> g $, respectively \cite{dutta2021quantum}. The dynamical map for a single-qubit system is:
\begin{equation}
    \begin{aligned}
      \epsilon^{NMAD}(\pmb{\rho}) = \mathbf{K}_0\pmb{\rho}\mathbf{K}^{\dag}_0 + \mathbf{K}_1\pmb{\rho}\mathbf{K}^{\dag}_1.  
      \end{aligned}\label{ADCfinalrho}
\end{equation}
A single-qubit system DWFs evolution under (non)-Markovian AD noise using the Bloch vector representation, given in Sec. \ref{single-qubit}, with  Eq. (\ref{ADCfinalrho}) and Eq. (\ref{DWFformula}) for a particular association of MUBs, can be seen to be 
\begin{widetext}
\begin{eqnarray}
\nonumber
W_{1, 1} &=& \frac{-(-1 + a_3)e^{-gt}(\gamma + (-g + \gamma)\cosh(l t) - l \sinh{(l t)}}{4(g - 2\gamma)}
+ \frac{1}{4}\left( 2 - (a_1 + a_2)T(g, \gamma, t)\right),\\\nonumber
W_{1, 2} &=& -\frac{1}{8}(-1 + a_3)e^{-gt}(1 + \cosh{(l t)})
- \frac{e^{-gt}\left( (-1 + a_3)g \sinh{(l t/2)}^2 + (-1 + a_3) l \sinh{(l t)} + (a_1 - a_2) e^{gt} (g - 2\gamma) T(g, \gamma, t) \right)}{4(g-2\gamma)},\\\nonumber
 W_{2, 1} &=& \frac{-(-1 + a_3)e^{-gt}(\gamma + (-g + \gamma)\cosh(l t) - l \sinh{(l t)}}{4(g - 2\gamma)}
+ \frac{1}{4}\left( 2 + (a_1 + a_2)T(g, \gamma, t)\right),\\
W_{2, 2} &=& -\frac{1}{8}(-1 + a_3)e^{-gt}(1 + \cosh{(l t)})- \frac{e^{-gt}\left( (-1 + a_3)g \sinh{(l t/2)}^2 + (-1 + a_3) l \sinh{(l t)} - (a_1 - a_2) e^{gt} (g - 2\gamma) T(g, \gamma, t) \right)}{4(g-2\gamma)},\nonumber\\
\end{eqnarray}
\end{widetext}
where, $T(g, \gamma, t) = \frac{\sqrt{e^{-g t}(-\gamma + (g - \gamma)\cosh{l t} + l\sinh{l t})}}{g - 2\gamma}$ and $a_1$, $a_2$, and $a_3$ are the componenets of the single-qubit Bloch vector $\textbf{a}$.
Plots of the DWFs for the $NS_1$ state of the single-qubit are shown in Fig. (\ref{qubitDWFNMAD}) and Fig. (\ref{qubitDWFMAD}) for non-Markovian and Markovian cases, respectively. The kinks at the peaks in the variation of $W_{2, 1}$ in Fig. (\ref{qubitDWFNMAD}) are due to the normalization property of the DWFs.
\begin{figure}[!htpb]
    \centering
    \includegraphics[height=85mm,width=1\columnwidth]{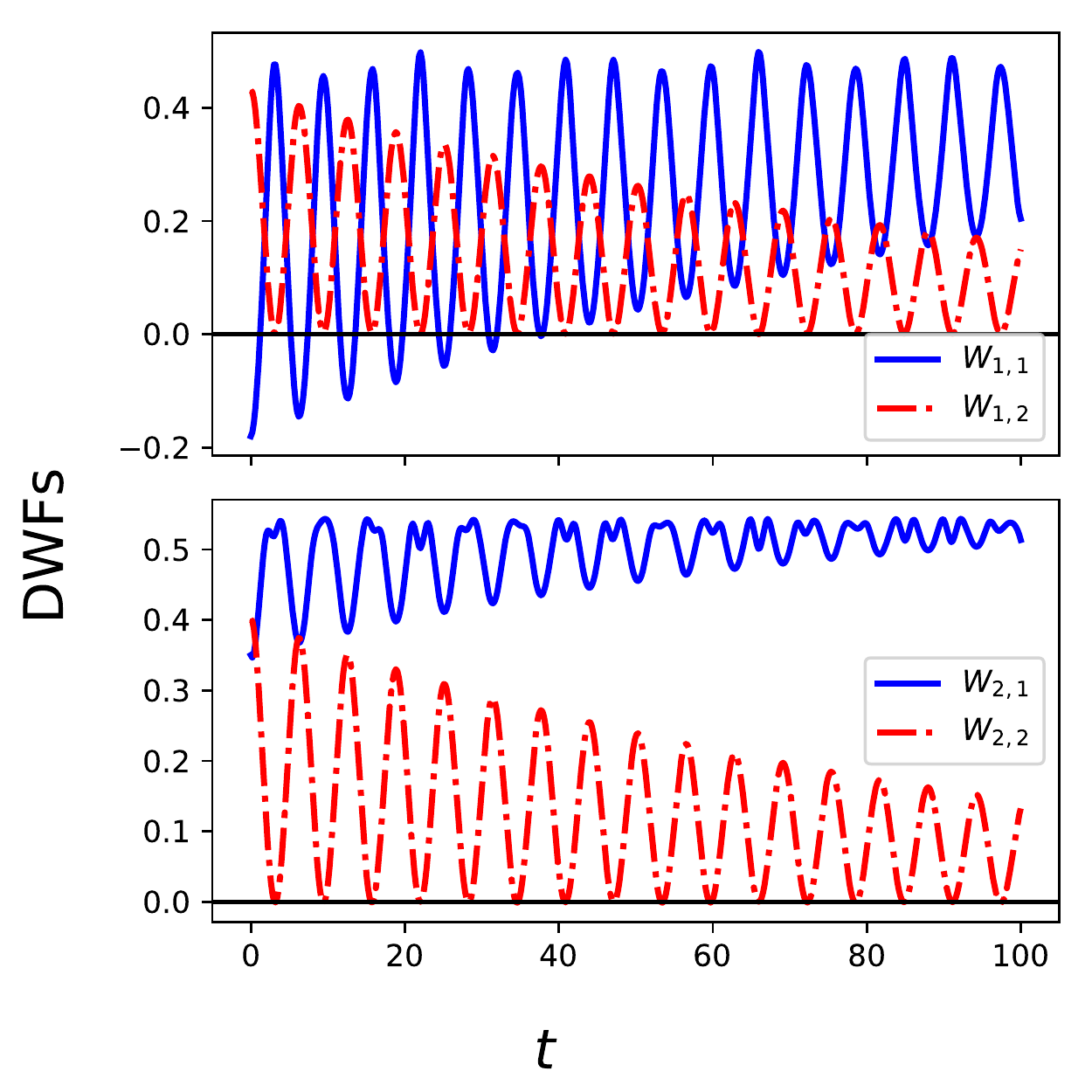}
    \caption{Variation of DWFs corresponding to the qubit's $NS_1$ state (for $a_1 = 0.50$, $a_2 = 0.56$, and $a_3 = -0.66$), under non-Markovian AD noise (for $\gamma = 50$, $g = 0.01$) with time.}
    \label{qubitDWFNMAD}
\end{figure}
\begin{figure}[!htpb]
    \centering
    \includegraphics[height=65mm,width=1\columnwidth]{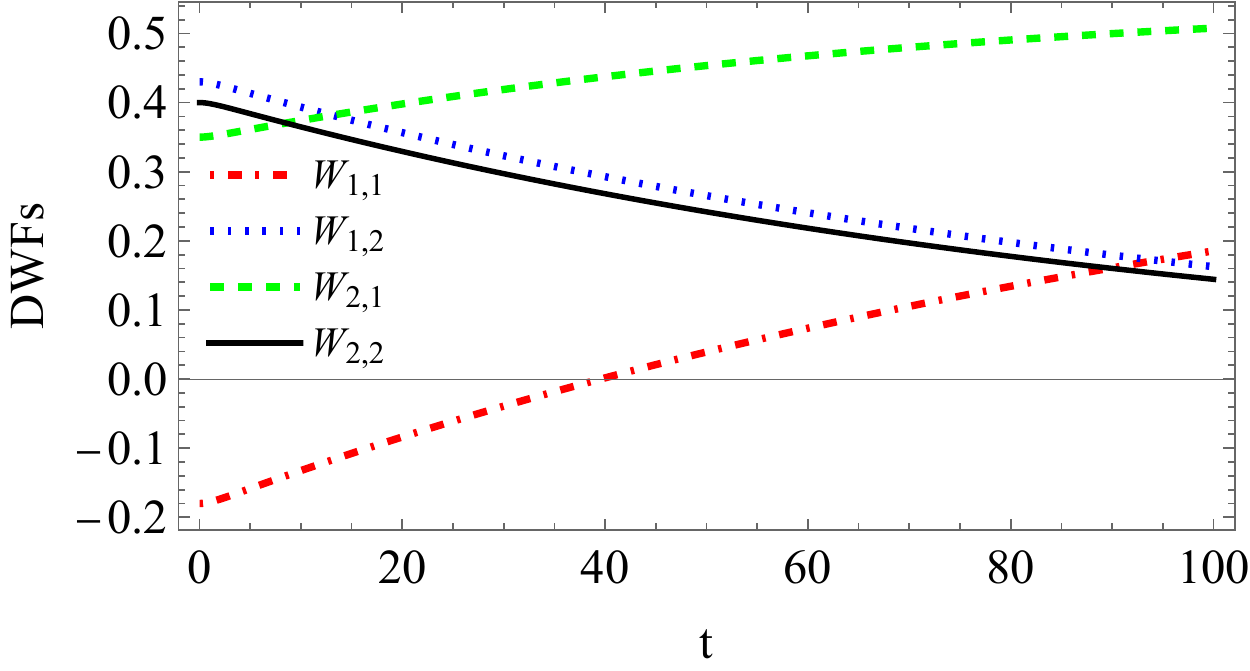}
    \caption{Variation of DWFs corresponding to the qubit's $NS_1$ state (for $a_1 = 0.50$, $a_2 = 0.56$, and $a_3 = -0.66$), under Markovian AD noise (for $\gamma = 0.01$, $g = 1$) with time.}
    \label{qubitDWFMAD}
\end{figure}
\subsection{\label{qutrit}Single-qutrit} 
A $3 \times 3$ real array defines the discrete phase space for single-qutrit systems. Elements of the Galois field $\mathcal{F}_3 = \{0, 1, \omega\}$ are used to label the points in this discrete phase space. This phase space has four possible striations, each of which possesses the predefined characteristics $(i)$, $(ii)$, and $(iii)$, as elaborated in Sec. \ref {stria} and shown in Fig. (\ref{striation2}). A set of four possible MUBs given in TABLE \ref{table2} are required for a one-to-one mapping with striations to calculate DWFs of single-qutrit systems \cite{brierley2009all}. 
The Bloch vector representation for a single-qutrit $\it, i.e.,$ $\pmb\rho = \frac{1}{3}(\textbf{I}_{3} + \sqrt{3} \textbf{n} . \bm{\lambda})$, (here, $\textbf{n} \in \textbf{R}^8$, $\textbf{I}_{3}$ is an identity operator and $\bm{\lambda}$'s are eight Gell-Mann matrices to describe a generalization of the Bloch ball representation of qubit to the case of qutrit given in \cite{goyal2016geometry} 
), and Eq. (\ref{DWFformula}) are used to determine the DWFs of qutrit for a particular association of MUBs given in TABLE \ref{table2} as
\begin{eqnarray}
      W_{1, 1} &=& \frac{1}{9} (1 + \sqrt{3} n_3 -\sqrt{3} n_6 - 3n_7 + n_8),\nonumber\\
      W_{1, 2} &=& \frac{1}{9} (1 - \sqrt{3} n_1 -\sqrt{3} n_3 - 3n_2 + n_8),\nonumber\\
     W_{1, 3} &=& \frac{1}{9} (1 - \sqrt{3} n_4 + 3n_5 - 2n_8),\nonumber\\
     W_{2, 1} &=& \frac{1}{9} (1 + \sqrt{3} n_3 -\sqrt{3} n_6 + 3n_7 + n_8),\nonumber\\
     W_{2, 2} &=& \frac{1}{9} (1 - \sqrt{3} n_1 -\sqrt{3} n_3 + 3n_2 + n_8),\nonumber\\
    W_{2, 3} &=& \frac{1}{9} (1 - \sqrt{3} n_4 - 3n_5 - 2n_8),\nonumber\\
     W_{3, 1} &=& \frac{1}{9} (1 + \sqrt{3} n_3 + 2\sqrt{3} n_6 + n_8),\nonumber\\
      W_{3, 2} &=& \frac{1}{9} (1 + 2\sqrt{3} n_1 - \sqrt{3} n_3 + n_8),\nonumber\\
     W_{3, 3} &=& \frac{1}{9} (1 + 2\sqrt{3} n_4 - 2n_8).
\end{eqnarray}
\begin{figure}[!htpb]
    \centering
    \includegraphics[width = 0.4\textwidth, height = 65mm]{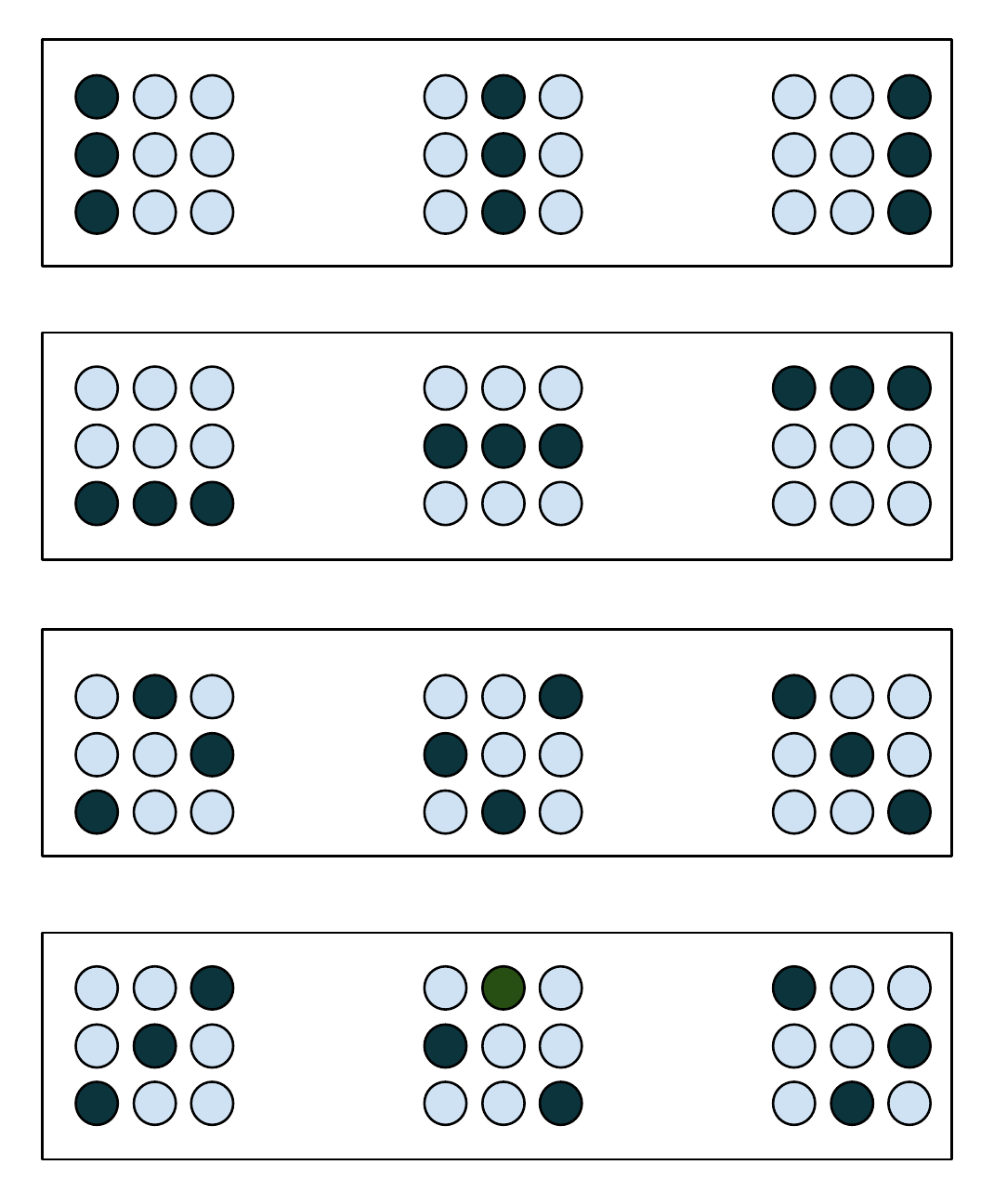}
    \caption{Lines and striations of the $3 \times 3$ phase space.}
    \label{striation2}
\end{figure}
\begin{table}
\begin{center}
\begin{tabular}{ | m{2cm}| m{5cm} | }

  \hline
  \textbf{Striation} & \textbf{MUBs associated with striation}\\ 
  \hline
  1 &  
  $\begin{pmatrix}
  1\\ 
  0\\
  0
\end{pmatrix}$,
$\begin{pmatrix}
  0\\ 
  1\\
  0
\end{pmatrix}$,
$\begin{pmatrix}
  0\\ 
  0\\
  1
\end{pmatrix}$\\
\hline
  2 &  
  $\frac{1}{\sqrt{3}}\begin{pmatrix}
  1\\ 
  1\\
  1
\end{pmatrix}$,
$\frac{1}{\sqrt{3}}\begin{pmatrix}
  1\\ 
  \omega\\
  \omega^2
\end{pmatrix}$,
$\frac{1}{\sqrt{3}}\begin{pmatrix}
  1\\ 
  \omega^2\\
  \omega
\end{pmatrix}$\\
\hline
  3 &  
  $\frac{1}{\sqrt{3}}\begin{pmatrix}
  1\\ 
  \omega^2\\
  \omega^2
\end{pmatrix}$,
$\frac{1}{\sqrt{3}}\begin{pmatrix}
  1\\ 
  1\\
  \omega
\end{pmatrix}$,
$\frac{1}{\sqrt{3}}\begin{pmatrix}
  1\\ 
  \omega\\
  1
\end{pmatrix}$\\
\hline
  4 &  
  $\frac{1}{\sqrt{3}}\begin{pmatrix}
  1\\ 
  \omega\\
  \omega
\end{pmatrix}$,
$\frac{1}{\sqrt{3}}\begin{pmatrix}
  1\\ 
  \omega^2\\
  1
\end{pmatrix}$,
$\frac{1}{\sqrt{3}}\begin{pmatrix}
  1\\ 
  1\\
  \omega^2
\end{pmatrix}$\\
\hline
\end{tabular}
\end{center}
\caption{\label{table2} The MUBs associated with lines of the $3 \times 3$ discrete phase space of single-qutrit systems. Here $\omega = e^{2\pi\iota/3}$ is a cube root of unity.}
\end{table}

\subsubsection{Random Telegraph noise}
For implementing the (non)-Markovian RTN channel dynamical map, the Kraus operators for a qutrit are given by \cite{daffer2004depolarizing, kumar2018non},
\begin{eqnarray}
      \mathbf{R}_0 = \sqrt{\frac{1 + \Lambda(t)}{2}}\textbf{I}_{3},
      \mathbf{R}_1 = \sqrt{\frac{1 - \Lambda(t)}{2}}\pmb{S_z},\nonumber\\
      \mathbf{R}_2 = \sqrt{\frac{1 - \Lambda(t)}{2}}\pmb{S},
\end{eqnarray}
where,
\begin{equation}
\pmb{S} = \frac{1}{2} \left(\pmb{S_x S_x} + \pmb{S_y S_y} - \pmb{S_z S_z}\right),
\end{equation}
and, 
\begin{eqnarray}
    \pmb{S_x} = \frac{1}{\sqrt{2}}\begin{pmatrix}
                               0 & 1 & 0\\
                               1 & 0 & 1\\
                               0 & 1 & 0
                               \end{pmatrix},
    \pmb{S_y} = \frac{1}{\sqrt{2}}\begin{pmatrix}
                               0 & -\iota & 0\\
                               \iota & 0 & -\iota\\
                               0 & \iota & 0
                               \end{pmatrix},\nonumber\\
    \pmb{S_z} =                 \begin{pmatrix}
                               1 & 0 & 0\\
                               0 & 0 & 0\\
                               0 & 0 & -1
                               \end{pmatrix},
     \textbf{I}_{3} =              \begin{pmatrix}
                               1 & 0 & 0\\
                               0 & 1 & 0\\
                               0 & 0 & 1
                               \end{pmatrix}.                         
\end{eqnarray}
The prior description of qubit's memory kernel $\Lambda(t)$ and criteria of (non)-Markovianty still holds, as in Sec. \ref{RTNqubit}. The dynamical map for a single-qutrit system is 
\begin{equation}
    \begin{aligned}
      \epsilon^{RTN}(\pmb{\rho}) = \mathbf{R}_0\pmb{\rho}\mathbf{R}^{\dag}_0 + \mathbf{R}_1\pmb{\rho}\mathbf{R}^{\dag}_1 + \mathbf{R}_2\pmb{\rho}\mathbf{R}^{\dag}_2.    
      \end{aligned}\label{RTNqutritfinalrho}
\end{equation}
The DWFs of the single-qutrit system are determined by first calculating its dynamical form $\epsilon^{RTN}(\pmb{\rho})$ using the Bloch vector representation of qutrit described in Sec. \ref{qutrit} and then using Eq. (\ref{DWFformula}) for a particular association of MUBs described in TABLE \ref{table2}. Fig. (\ref{qutritDWFNMRTN}) shows the variation of DWFs of the qutrit's $NS_1$ state with time for non-Markovian RTN. In the Markovian RTN regime, non-oscillatory behaviour is seen in contrast to the non-Markovian RTN case.
\begin{figure}[!htpb]
    \centering
    \includegraphics[height=65mm,width=1\columnwidth]{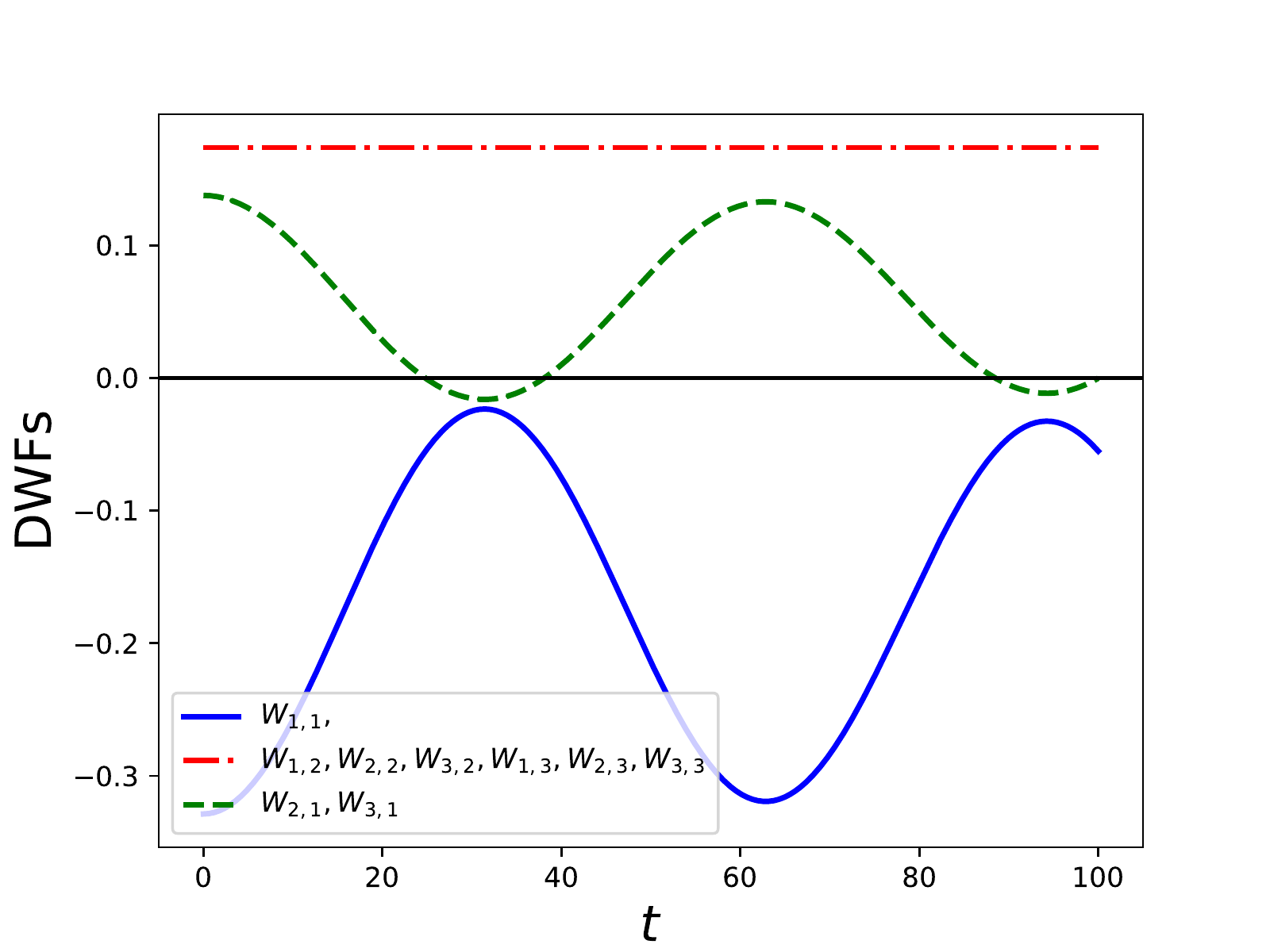}
    \caption{Variation of DWFs corresponding to the qutrit's $NS_1$ state (when $n_1 = 0$, $n_2 = 0$, $n_3 = -0.5$, $n_4 = 0$, $n_5 = 0$, $n_6 = 0.4$, $n_7 = 0.7$, $n_8 = -0.3$), under (non)-Markovian RTN (for $\gamma = 0.001$, $b = 0.05$) with time.}
    \label{qutritDWFNMRTN}
\end{figure}
\subsubsection{Amplitude Damping Noise}
The Kraus operators that define the (non)-Markovian AD channel for qutrits are \cite{ghosal2021characterizing, utagi2020ping,dutta2021quantum}
\begin{eqnarray}
\nonumber
    \textbf{K}_{0} &=& \begin{pmatrix}
     1 & 0 & 0\\
     0 & \sqrt{1 - \lambda(t)} & 0\\
     0 & 0 & \sqrt{1 - \lambda(t)}
    \end{pmatrix},\\
    \textbf{K}_{1} &=& \begin{pmatrix}
    0 & \sqrt{\lambda(t)} & 0\\
    0 & 0 & 0\\
    0 & 0 & 0
    \end{pmatrix},\nonumber\\
    \textbf{K}_{2} &=& \begin{pmatrix}
    0 & 0 & \sqrt{\lambda(t)}\\
    0 & 0 & 0\\
    0 & 0 & 0
\end{pmatrix}.
\end{eqnarray}
The dynamical map form is
\begin{equation}
    \begin{aligned}
      \epsilon^{NMAD}(\pmb{\rho}) = \mathbf{K}_0\pmb{\rho}\mathbf{K}^{\dag}_0 + \mathbf{K}_1\pmb{\rho}\mathbf{K}^{\dag}_1 + \mathbf{K}_2\pmb{\rho}\mathbf{K}^{\dag}_2.    
      \end{aligned}\label{NMADqutritfinalrho}
\end{equation}
The expression for $\lambda(t)$ and the regimes of (non)-Markovian behaviour remain unchanged, as in Sec. \ref{qubitNMAD}. To study the behaviour of single-qutrit DWFs under the (non)-Markovian AD channel, we use its dynamical map Eq. (\ref{NMADqutritfinalrho}) and (\ref{DWFformula}) for a particular association of MUBs given in TABLE \ref{table2}. For the qutrit's $NS_1$ state, Fig. (\ref{qutritDWFNMAD}) display the DWFs variation for the non-Markovian AD case. Compared to the single-qubit's $NS_1$ state DWFs, single-qutrit's $NS_1$ state DWFs have a higher negative value and remain negative for a longer time, as depicted by Fig. (\ref{qutritDWFNMAD}).
\begin{figure}[!htpb]
    \centering
    \includegraphics[height=65mm,width=1\columnwidth]{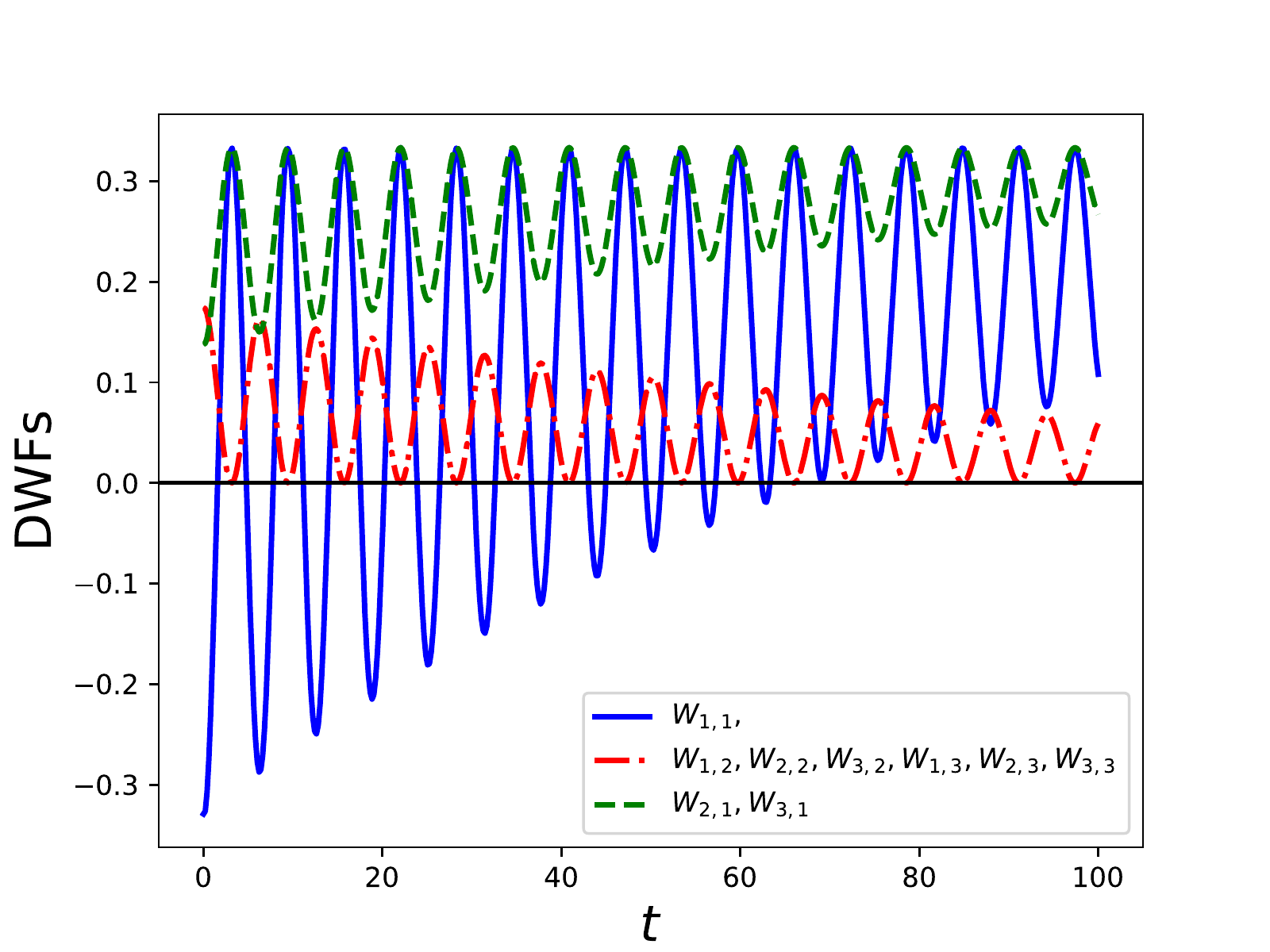}
    \caption{Variation of DWFs corresponding to the the qutrit's $NS_1$ state (for $n_1 = 0$, $n_2 = 0$, $n_3 = -0.5$, $n_4 = 0$, $n_5 = 0$, $n_6 = 0.4$, $n_7 = 0.7$, $n_8 = -0.3$), under non-Markovian AD (for $\gamma = 50$, $g = 0.01$) with time.}
    \label{qutritDWFNMAD}
\end{figure}
\subsection{\label{two-qubit}Two-qubit} 
The discrete phase space for two-qubit systems is defined on a $4 \times 4$ array. The Galois field, $\mathcal{F}_4 = \{0, 1, \omega, \omega^2\}$ elements are used to label the points in this discrete phase space. There are five possible sets of parallel lines (striations), each of which satisfies the (i), (ii), and (iii) properties listed in Sec. \ref {stria} and depicted by Fig. (\ref{striation3}). A set of five MUBs are needed for a one-to-one mapping with striations, as discussed in \cite{durt2010mutually}, and is provided in TABLE \ref{table3}. A two-qubit system is represented as: 
\begin{equation}
\pmb{\rho} = \frac{1}{4}(\pmb{I_{2}} \otimes \pmb{I_{2}} + \sum_{i = 1} ^{3} a_{i} \pmb{\sigma_{i}} \otimes \pmb{I_{2}} + \sum_{i = 1} ^{3} s_{i} \pmb{I_{2}} \otimes \pmb{\sigma_{i}}  + \sum_{i,j = 1} ^{3} t_{ij} (\pmb{\sigma_{i}} \otimes \pmb{\sigma_{j}})),
\label{2-qubit-rho}
\end{equation}
where $\pmb{I_{2}}$ denotes identity operator, $\pmb{\sigma_{i}}$'s denote the standard Pauli matrices, $a_{i}$'s and $s_{i}$'s are components of vectors in $\textbf{R}^{3}$. The coefficients $t_{ij} = Tr(\pmb{\rho} \pmb{\sigma_{i}} \otimes \pmb{\sigma_{j}})$ combine to give a real matrix $\textbf{T}$, known as the correlation matrix.
We use Eq. (\ref{2-qubit-rho}) in  Eq. (\ref{DWFformula}) to find the DWFs of the two-qubit systems, which are shown in Appendix A.
\begin{figure}[!htpb]
    \centering
    \includegraphics[width = 0.4\textwidth, height = 65mm]{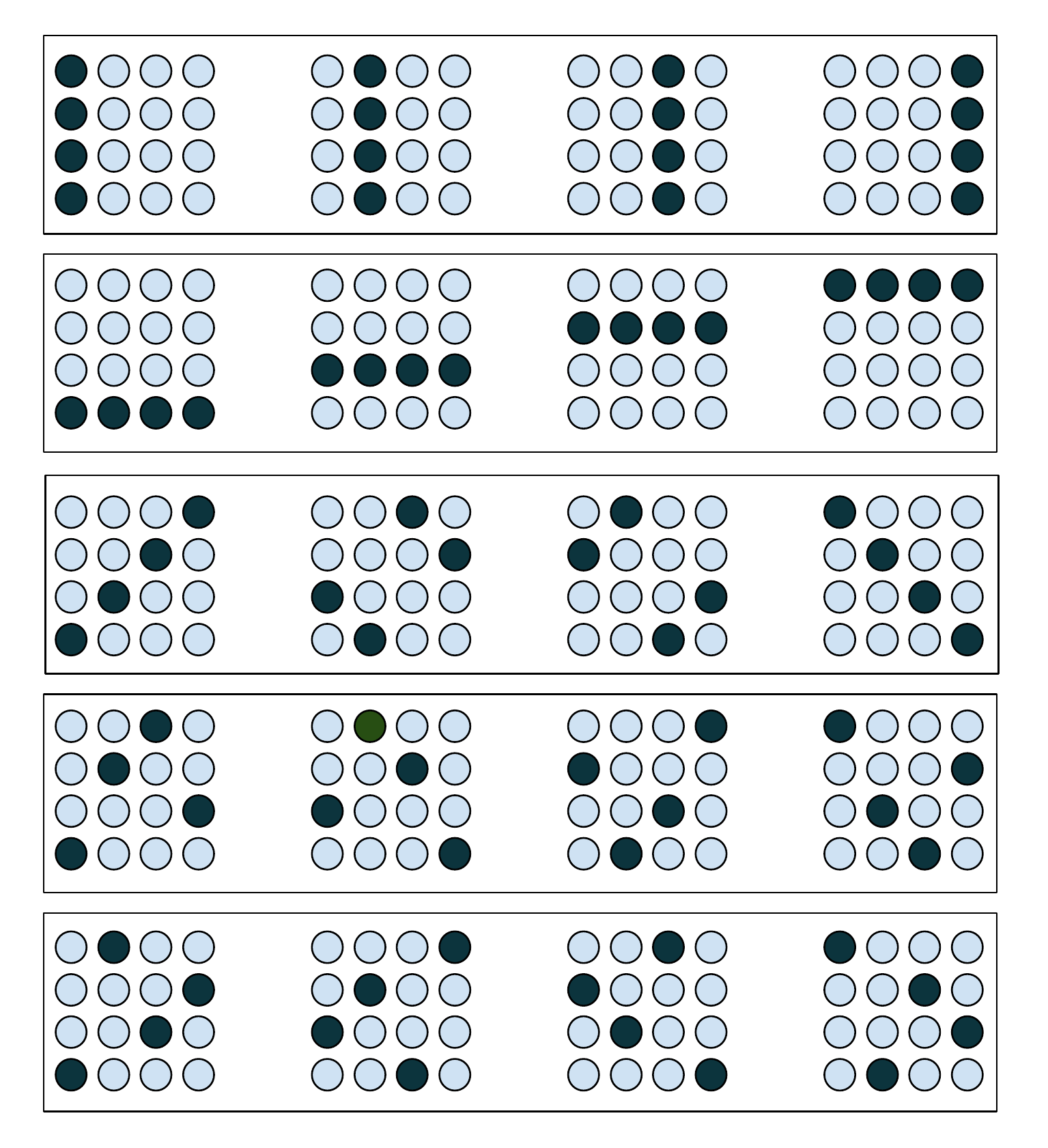}
    \caption{Lines and striations of the $4 \times 4$ phase space.}
    \label{striation3}
\end{figure}
\begin{table}
\begin{center}
\begin{tabular}{ | m{2cm}| m{6cm} | }

  \hline
  \textbf{Striation} & \textbf{MUBs associated with striation}\\ 
  \hline
  1 &  
  $\begin{pmatrix}
  1\\ 
  0\\
  0\\
  0
\end{pmatrix}$,
$\begin{pmatrix}
  0\\ 
  1\\
  0\\
  0
\end{pmatrix}$,
$\begin{pmatrix}
  0\\ 
  0\\
  1\\
  0
\end{pmatrix}$,
$\begin{pmatrix}
  0\\ 
  0\\
  0\\
  1
\end{pmatrix}$\\
\hline
  2 &  
  $\frac{1}{2}\begin{pmatrix}
  1\\ 
  1\\
  1\\
  1
\end{pmatrix}$,
$\frac{1}{2}\begin{pmatrix}
  1\\ 
 -1\\
  1\\
  -1
\end{pmatrix}$,
$\frac{1}{2}\begin{pmatrix}
  1\\ 
  1\\
  -1\\
  -1
\end{pmatrix}$,
$\frac{1}{2}\begin{pmatrix}
  1\\ 
  -1\\
  -1\\
  1
\end{pmatrix}$\\
\hline
  3 &  
$\frac{1}{2}\begin{pmatrix}
  1\\ 
  -\iota\\
   \iota\\
   1
\end{pmatrix}$,
$\frac{1}{2}\begin{pmatrix}
  1\\ 
  \iota\\
   \iota\\
   -1
\end{pmatrix}$,
$\frac{1}{2}\begin{pmatrix}
  1\\ 
  -\iota\\
  -\iota\\
   -1
\end{pmatrix}$,
$\frac{1}{2}\begin{pmatrix}
  1\\ 
  \iota\\
  -\iota\\
   1
\end{pmatrix}$\\
\hline
  4 &  
$\frac{1}{2}\begin{pmatrix}
  1\\ 
  1\\
  \iota\\
   -\iota
\end{pmatrix}$,
$\frac{1}{2}\begin{pmatrix}
  1\\ 
  -1\\
  \iota \\
   \iota
\end{pmatrix}$,
$\frac{1}{2}\begin{pmatrix}
  1\\ 
  1\\
  -\iota\\
   \iota
\end{pmatrix}$,
$\frac{1}{2}\begin{pmatrix}
  1\\ 
  -1\\
  -\iota\\
   -\iota
\end{pmatrix}$\\
\hline
  5 &  
$\frac{1}{2}\begin{pmatrix}
  1\\ 
  -\iota\\
   1\\
  \iota
\end{pmatrix}$,
$\frac{1}{2}\begin{pmatrix}
  1\\ 
  \iota\\
   1\\
   -\iota
\end{pmatrix}$,
$\frac{1}{2}\begin{pmatrix}
  1\\ 
  -\iota\\
  -\iota\\
   -\iota\\
\end{pmatrix}$,
$\frac{1}{2}\begin{pmatrix}
  1\\ 
  \iota\\
  -1\\
   \iota
\end{pmatrix}$\\
\hline
\end{tabular}
\end{center}
\caption{\label{table3} The MUBs associated with lines of the $4 \times 4$ discrete phase space of two-qubit systems.}
\end{table}
\subsubsection{Random Telegraph Noise}
The dynamical map of the local interaction of two qubits with (non)-Markovian RTN channel is 
\begin{equation}
   \begin{aligned}
      \epsilon^{RTN}(\pmb{\rho}) = \sum_{i = 0}^{1}\sum_{j = 0}^{1} (\mathbf{R}_i \otimes \mathbf{R}_j) \pmb{\rho} (\mathbf{R}_i \otimes \mathbf{R}_j)^{\dag}.
      \end{aligned}\label{2qubitRTNfinalrho}
\end{equation}
Here, $\mathbf{R}_i$ and $\mathbf{R}_j$ are as discussed for the qubit case above in Sec. \ref{RTNqubit}. The DWFs of two qubits can be constructed using Eq. (\ref{2qubitRTNfinalrho}) and Eq.(\ref{DWFformula}) for a particular association of MUBs given in TABLE \ref{table3}. Figure (\ref{2qubitDWFNMRTN}) depicts how the non-Markovian RTN scenario's two-qubit $NS_1$ state changes over time.
\begin{figure}[!htpb]
    \centering
    \includegraphics[height=85mm,width=1\columnwidth]{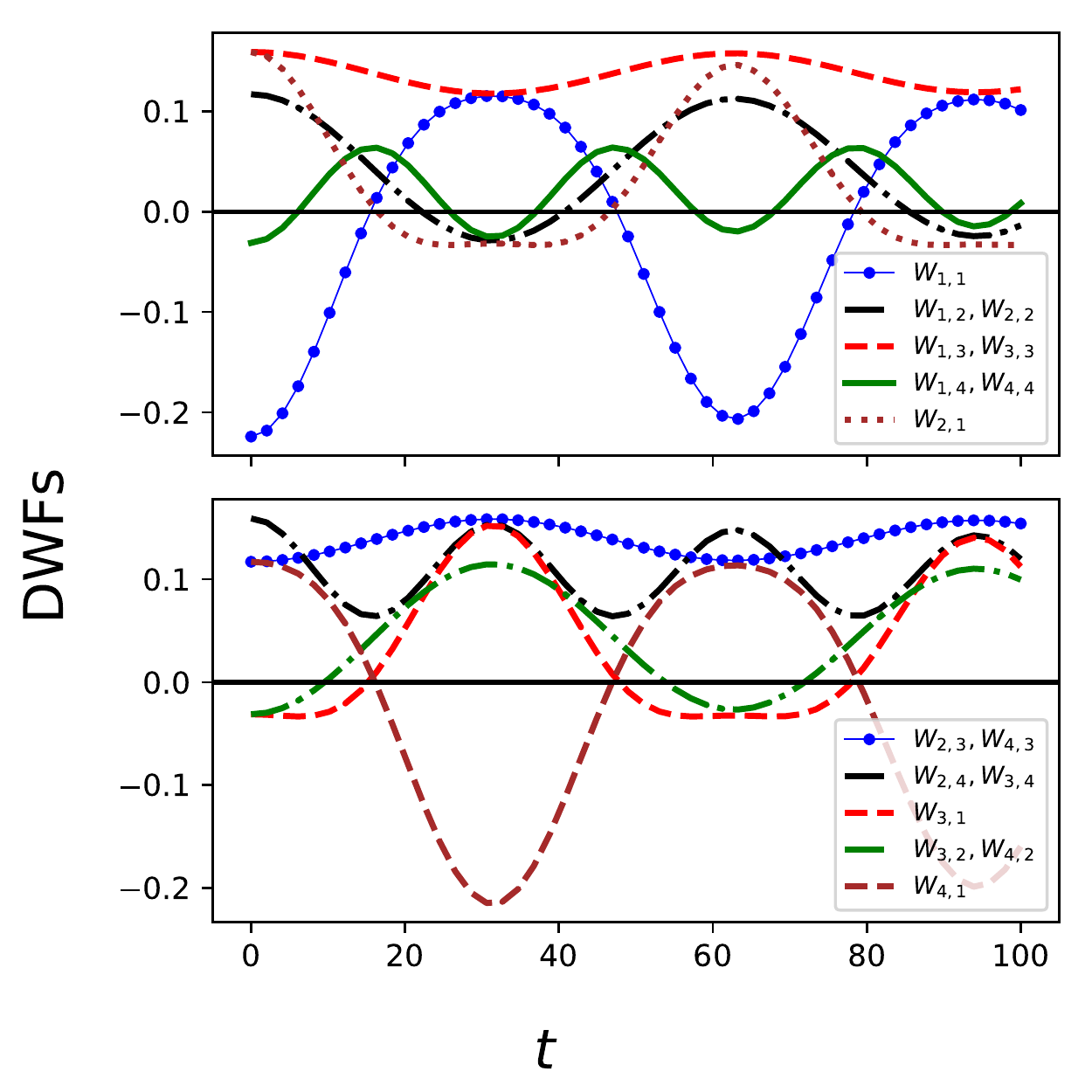}
    \caption{Variation of DWFs corresponding to the two-qubit's $NS_1$ state ($a_1 = 0.14$, $a_2 = 0.14$, $a_3 = 0.61$, $s_1 = 0.44$, $s_2 = -0.44$, $s_3 = 0.14$, $t_{11} = 0.61$, $t_{12} = 0.14$, $t_{13} = -0.44$, $t_{21} = -0.14$, $t_{22} = -0.61$, $t_{23} = -0.44$, $t_{31} = 0.61$, $t_{32} = -0.61$ and, $t_{33} = 0.44$), under non-Markovian RTN (for $\gamma = 0.001$ and $b = 0.05$).}
    \label{2qubitDWFNMRTN}
\end{figure}
\subsubsection{Amplitude Damping Noise}
The dynamical map for the local interaction of two-qubit systems with (non)-Markovian AD channel acts as
\begin{equation}
   \begin{aligned}
      \epsilon^{NMAD}(\pmb{\rho}) = \sum_{i = 0}^{1}\sum_{j = 0}^{1} (\mathbf{K}_i \otimes \mathbf{K}_j) \pmb{\rho} (\mathbf{K}_i \otimes \mathbf{K}_j)^{\dag}.
      \end{aligned}\label{2qubitNMADfinalrho}
\end{equation}
The Kraus operators, $\textbf{K}_{0}$ and $\textbf{K}_{1}$, are as defined for the qubit's (non)-Markovian AD. Using its dynamical form Eq. (\ref{2qubitNMADfinalrho}) and Eq. (\ref{DWFformula}) for a particular association of MUBs given in TABLE \ref{table3}, we have built its DWFs. Figure (\ref{2qubitDWFNMAD}) illustrates the behaviour of the two-qubit $NS_1$ state under the non-Markovian AD evolution. Compared to single-qubit's $NS_1$ state DWFs, two-qubit's $NS_1$ state DWFs have a higher negative value. It is less than the single-qutrit case but sustains negative values for much longer than the single qubit and qutrit and is shown in Fig. (\ref{2qubitDWFNMAD}).   
\begin{figure}[!htpb]
    \centering
    \includegraphics[height=85mm,width=1\columnwidth]{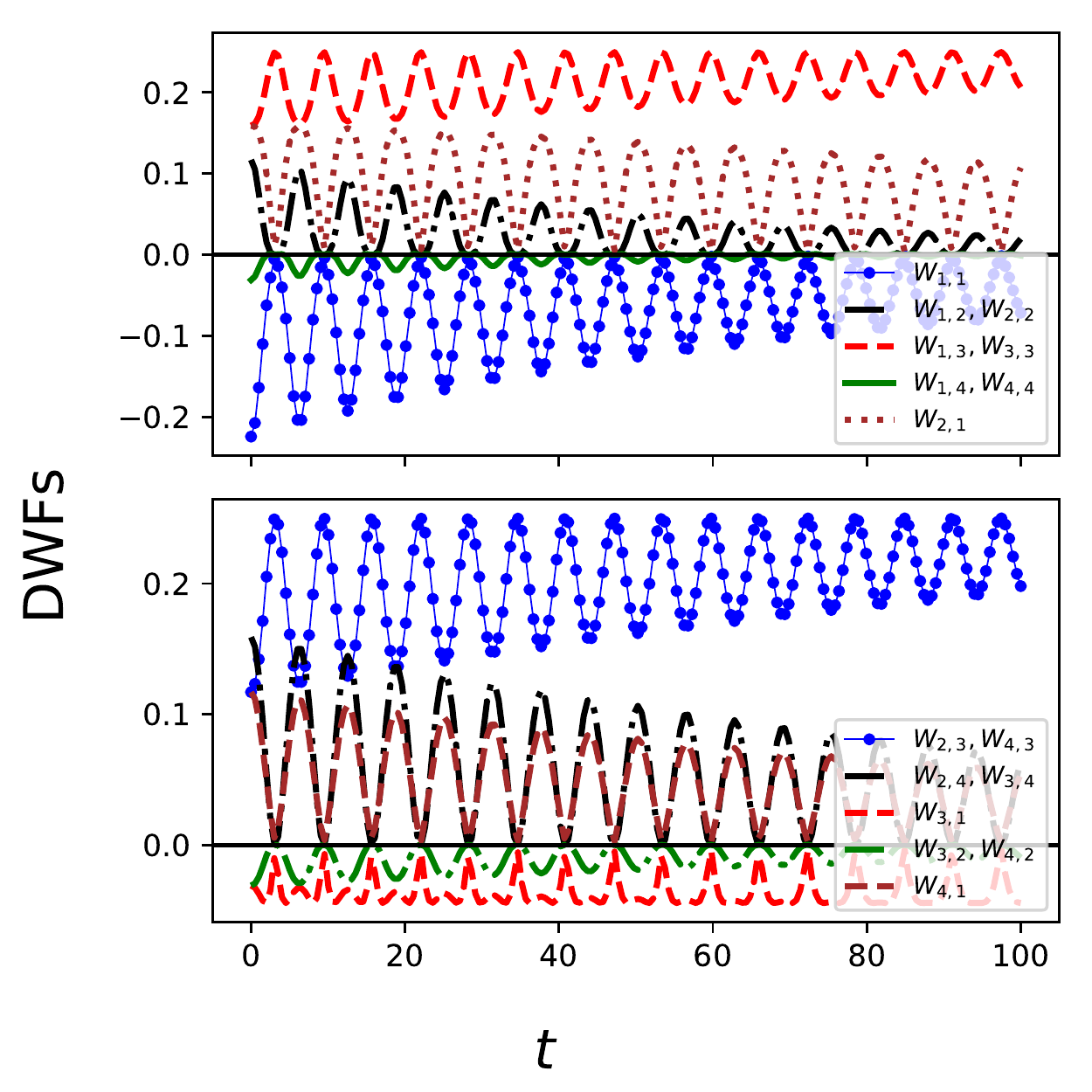}
    \caption{Variation of DWFs corresponding to the two-qubit's $NS_1$ state ($a_1 = 0.14$, $a_2 = 0.14$, $a_3 = 0.61$, $s_1 = 0.44$, $s_2 = -0.44$, $s_3 = 0.14$, $t_{11} = 0.61$, $t_{12} = 0.14$, $t_{13} = -0.44$, $t_{21} = -0.14$, $t_{22} = -0.61$, $t_{23} = -0.44$, $t_{31} = 0.61$, $t_{32} = -0.61$ and, $t_{33} = 0.44$), under (non)-Markovian AD (for $\gamma = 50$, $g = 0.01$) with time.}
    \label{2qubitDWFNMAD}
\end{figure}
\section{\label{neg}Discrete Wigner negativity and mana variation  under different noisy channels}
Discrete Wigner negativity $|N_G(\pmb{\rho})|$, (for $d = 2, 3, 4$), and mana, (for $d = 3$) are studied next under (non)-Markovian noisy channels using Eq. (\ref{negativity}) and Eq. (\ref{mana}), respectively. Figures (\ref{negativityNMRTN}) and (\ref{negativityNMAD}) show how $|N_G(\pmb{\rho})|$ changes when subjected to several noisy models, including non-Markovian RTN and non-Markovian AD respectively. Discrete Wigner negativity is highest for qutrit compared to qubit and two-qubit. Under the influence of (non)-Markovian AD noise, it falls rapidly compared to the two qubits and sustains for a longer duration than the single qubit, which is depicted by Fig. (\ref{negativityNMAD}). Under non-Markovian RTN, all 
the cases {\it, i.e.,} qubit, qutrit, two-qubit, show expected oscillatory behavior, with the peaks and dips of qubit and two-qubit in synchronization, with an alternate pattern with the qutrit, see Fig. (\ref{negativityNMRTN}). Figure (\ref{mana-qutrit-NMAD}) displays how mana varies for a qutrit's $NS_1$ and $NS_2$ state when subjected to non-Markovian AD and RTN noise. As we can see from Fig. (\ref{mana-qutrit-NMAD}), initially, the $NS_1$ state has a higher value of mana than the $NS_2$ state. However, it dies off very quickly in comparison to the $NS_2$ state. Hence, the $NS_2$ state persists longer and has a finite mana value. Under the non-Markovian RTN, mana for both the negative quantum states of qutrit show expected oscillatory behaviour which is persistent for much longer than the non-Markovian AD. Interestingly, the action of the phase S-gate \cite{li2023optimal} produces  the conjugate of both the negative quantum states of the qutrit.
\begin{figure}[!htpb]
    \centering
    \includegraphics[height=55mm,width=1\columnwidth]{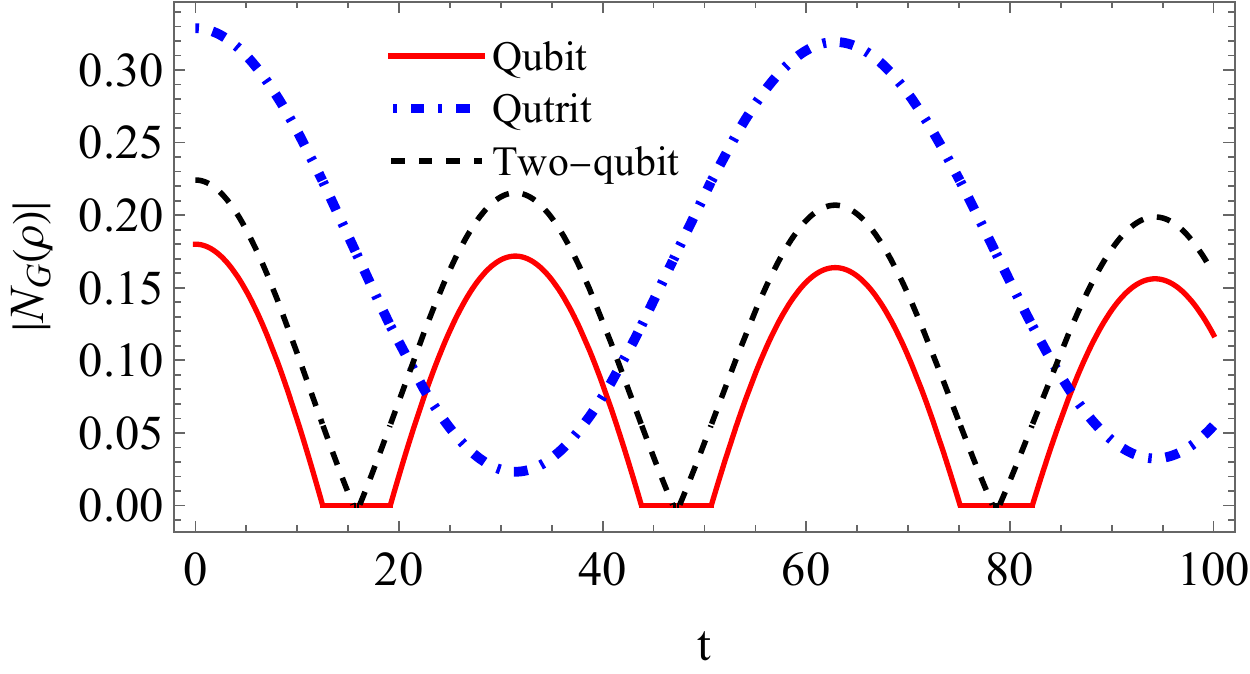}
    \caption{Variation of discrete Wigner negativity for a qubit, qutrit, and two-qubit systems non-Markovian RTN with time. For $\gamma = 0.001$ and $b = 0.05$.}
    \label{negativityNMRTN}
\end{figure}
\begin{figure}[!htpb]
    \centering
    \includegraphics[height=55mm,width=1\columnwidth]{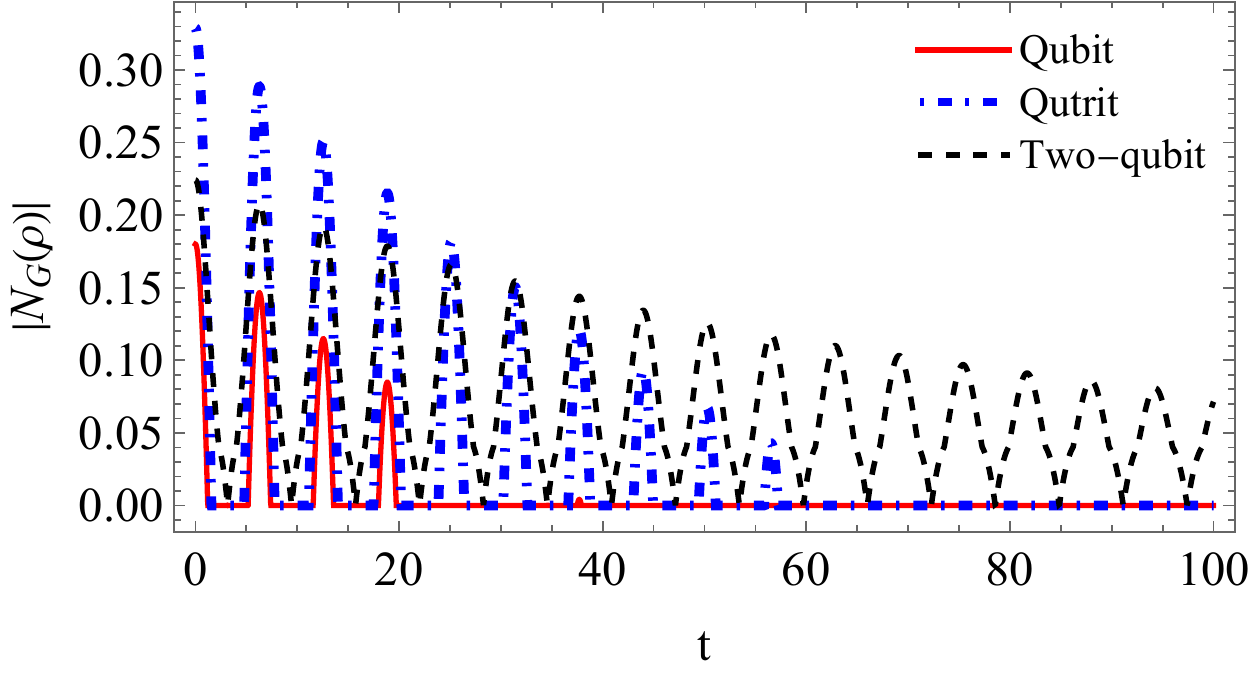}
    \caption{Variation of discrete Wigner negativity for a qubit, qutrit, and two-qubit systems under non-Markovian AD noise with time. For $\gamma = 50$, $g = 0.01$.}
    \label{negativityNMAD}
\end{figure}
\begin{figure}[!htpb]
    \centering
    \includegraphics[height=65mm,width=1\columnwidth]{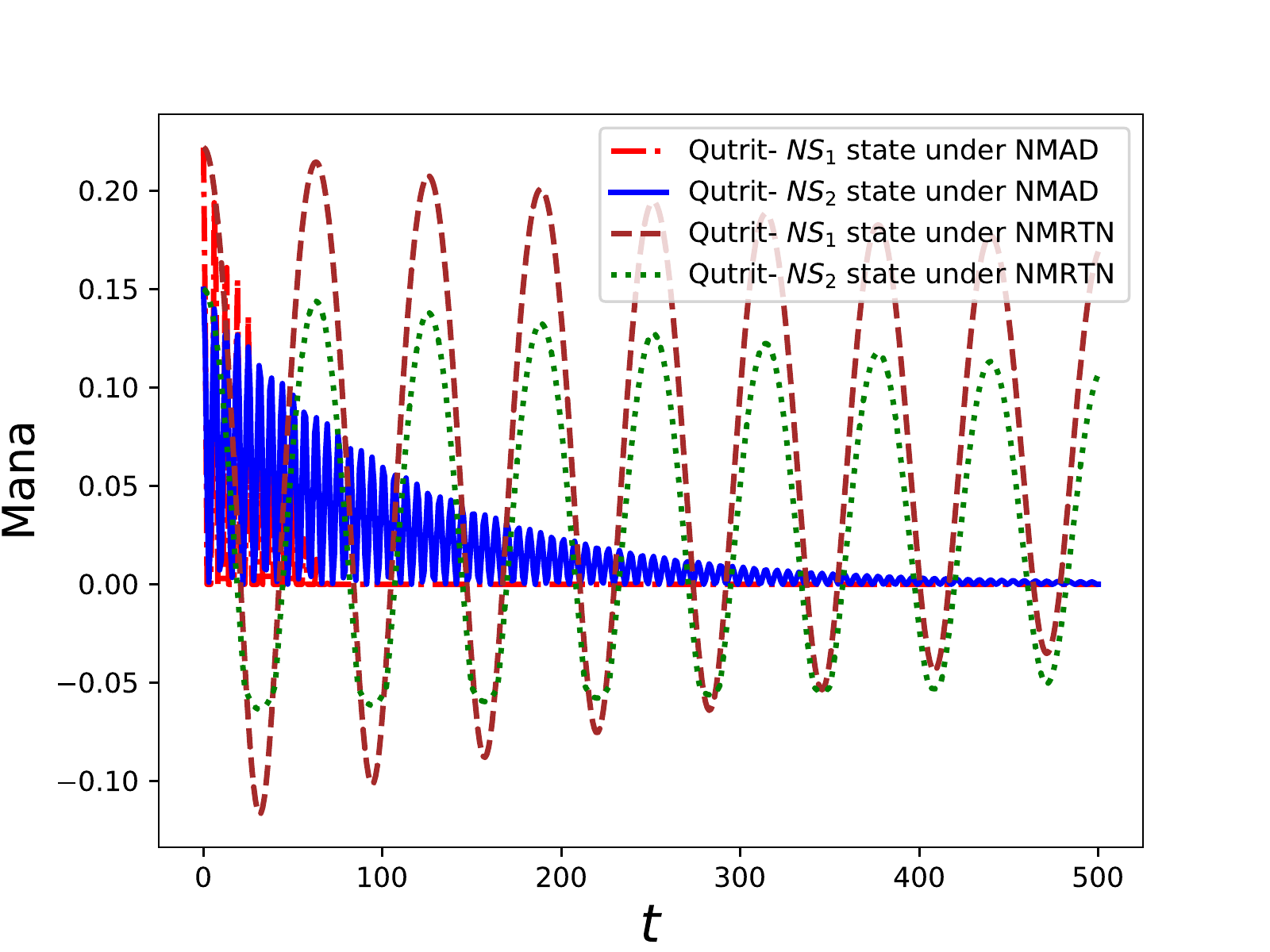}
    \caption{Variation of mana of a qutrit's $NS_1$ and $NS_2$ state under non-Markovian AD ($\gamma = 50$, $g = 0.01$) and non-Markovian RTN ($\gamma = 0.001$, $b = 0.05$) noise with time.}
    \label{mana-qutrit-NMAD}
\end{figure}

\section{\label{coh, con}Quantum coherence and entanglement using DWFs under different noisy channels: two-qubit systems} 
A number of methods for determining a quantum system's coherence are available in the literature \cite{baumgratz2014quantifying, girolami2014observable}. We particularly focus here on the $l_{1}$ norm of coherence, defined as the sum of the absolute values of all off-diagonal elements of $\pmb{\rho}$ as given below \cite{baumgratz2014quantifying}.
\begin{equation}
   \begin{aligned}
     C_{l_{1}}(\pmb{\rho}) = \sum_{i \neq j} |\pmb{\rho}_{i,j}|.
      \end{aligned}
      \label{coherence_Eq.}
\end{equation}
The variation in quantum coherence for the two-qubit $NS_1$, $NS_2$, and $NS_3$ states as well as Bell states under the operation of (non)-Markovian RTN and AD channels is next examined using Eq. (\ref{rho-decomposition-in-A}), $\textit{i.e.}$, the DWFs form of the states, and the above Eq. (\ref{coherence_Eq.}). From Figs. (\ref{coherence_NMRTN}) and (\ref{coherence_NMAD}), it is clear that the two-qubit negative quantum states, $NS_1$, $NS_2$, and $NS_3$ have quantum coherence greater than the Bell state under non-Markovian RTN and AD noise channels. Initially, the $NS_{3}$ state has maximum quantum coherence in comparison to the $NS_{1}$, $NS_{2}$, and Bell state, as can be seen from Figs. (\ref{coherence_NMRTN}) and (\ref{coherence_NMAD}). Moreover, all the states display anticipated decaying oscillatory behaviour under the non-Markovian RTN and AD noise channels, as illustrated in Figs. (\ref{coherence_NMRTN}) and (\ref{coherence_NMAD}). Additionally, the quantum coherence of the $NS_1$ and $NS_2$ states under a non-Markovian AD noise channel is sustained for longer, as can be seen from Fig. (\ref{coherence_NMAD}).
\begin{figure}[!htpb]
    \centering
    \includegraphics[height=65mm,width=1\columnwidth]{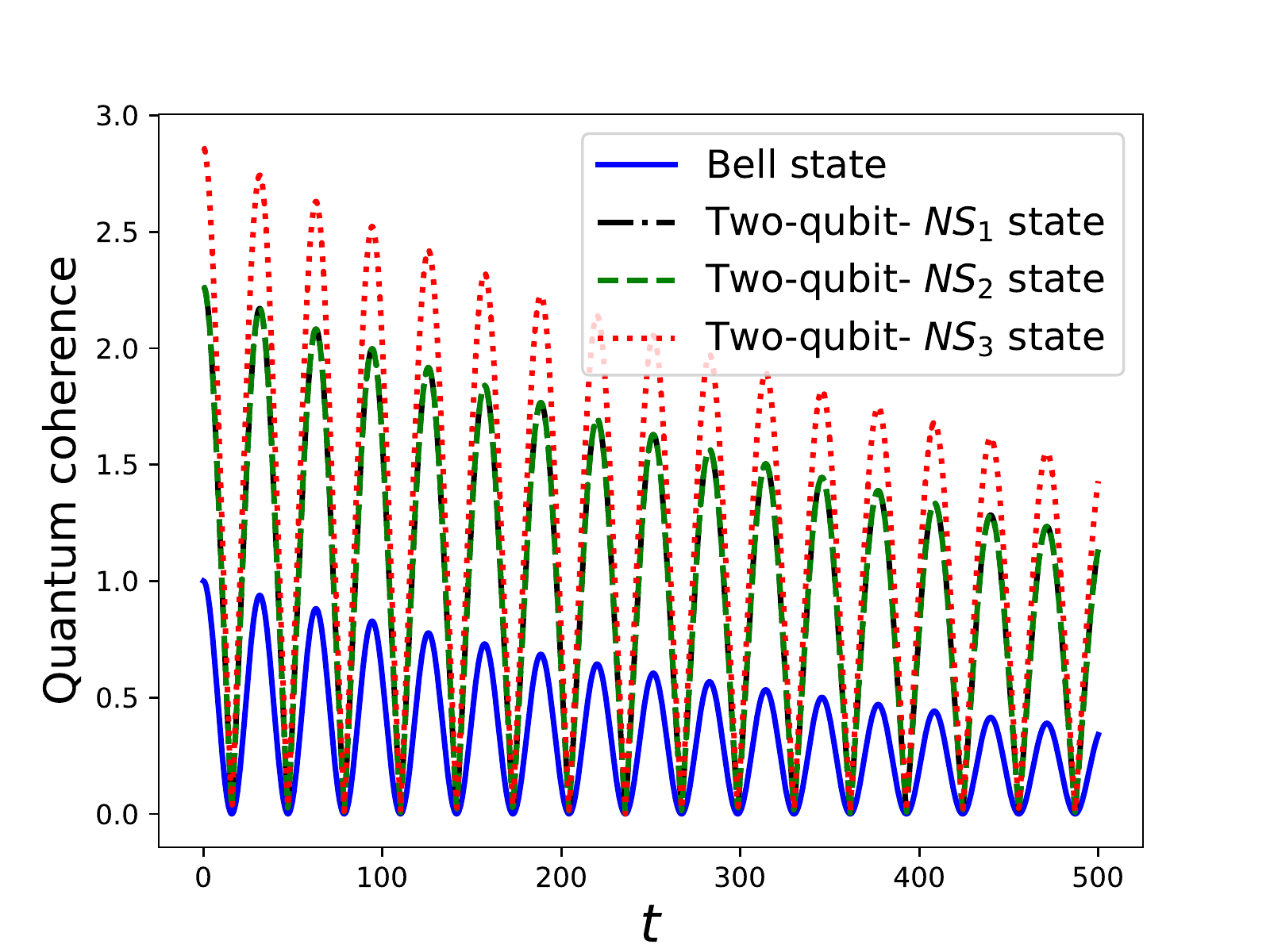}
    \caption{Variation of quantum coherence for the two-qubit's $NS_1$, $NS_2$, $NS_3$ states, and Bell state under non-Markovian RTN noise with time. For $\gamma = 0.001$ and $b = 0.05$.}
    \label{coherence_NMRTN}
\end{figure}
\begin{figure}[!htpb]
    \centering
    \includegraphics[height=65mm,width=1\columnwidth]{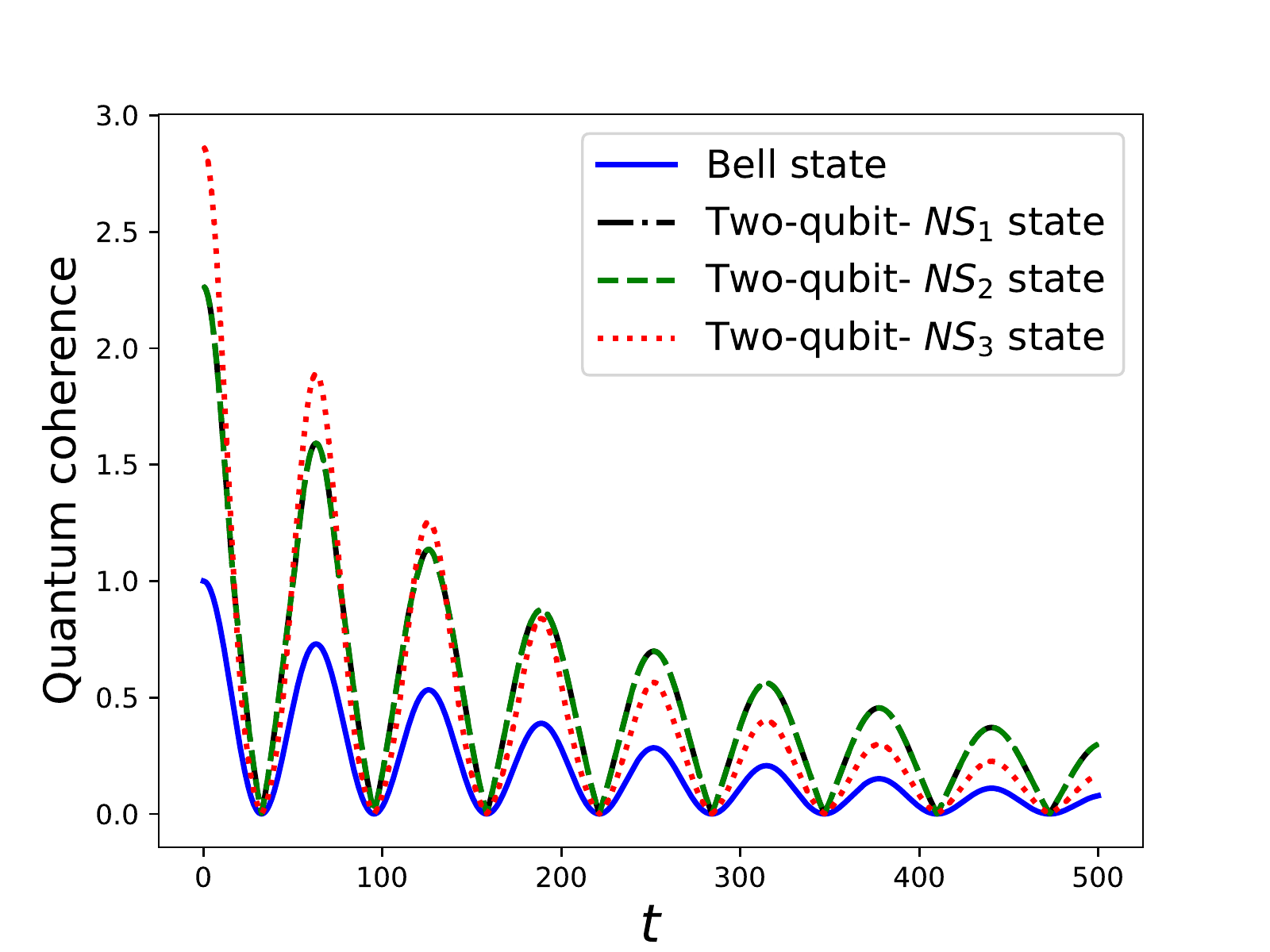}
    \caption{Variation of quantum coherence for the two-qubit's $NS_1$, $NS_2$, $NS_3$ states, and Bell state under non-Markovian AD noise with time. For $\gamma = 1$, $g = 0.005$.}
    \label{coherence_NMAD}
\end{figure}

Now we examine the variation of entanglement under (non)-Markovian noisy channels as it is one of the most crucial quantum information sources. For a two-qubit system, concurrence is an entanglement metric \cite{wootters1998entanglement}, which is defined as
\begin{equation}
   \begin{aligned}
     C(\pmb{\rho}_{AB}) = max \{0, \lambda_{1} - \lambda_{2} - \lambda_{3} - \lambda_{4}\},
      \end{aligned}
\end{equation}
Here $\lambda_{i}$'s are the eigenvalues of $\sqrt{\sqrt{\pmb{\rho}_{AB}} \tilde{\pmb{\rho}}_{AB} \sqrt{\pmb{\rho}_{AB}}}$ in the descending order and $\Tilde{\pmb{\rho}}_{AB} = (\sigma_{y} \otimes \sigma_{y}) \pmb{\rho}_{AB}^{*} (\sigma_{y} \otimes \sigma_{y})$, $\pmb{\rho}_{AB}^{*}$ is the complex conjugate of $\pmb{\rho}_{AB}$. We have $C(\pmb{\rho}_{AB}) = 0$ for separable states and $0 < C(\pmb{\rho}_{AB}) \leq 1$ for entangled states. Using the DWFs form of the states, Eq. (\ref{rho-decomposition-in-A}), we next analyze the variation in concurrence for the two-qubit's $NS_1$, $NS_2$, and $NS_3$ states and Bell states under the operation of (non)-Markovian RTN and AD channels. We can see from Figs. (\ref{concurNMRTN}) and (\ref{concurNMAD}) that at $t = 0$, the $NS_1$, $NS_2$, and $NS_3$ states have concurrence between zero and one, $\textit{i.e.}$, these states are entangled, an indicator of quantumness.  
Figure (\ref{concurNMRTN}) shows that under the action of a non-Markovian RTN channel, concurrence for all, $\textit{i.e.}$, $NS_1$, $NS_2$, and $NS_3$ states and Bell state exhibit decaying oscillations in synchronization with one another. Furthermore, from Fig. (\ref{concurNMAD}), we can see that the $NS_3$ state initially possesses concurrence equivalent to the Bell state. Still, with time, it attains a higher value of concurrence than the Bell state under non-Markovian AD noise. The $NS_1$ and $NS_2$ states are seen to have less concurrence in the beginning, but with the passage of time, they also attain higher concurrence in comparison to the Bell states. 
\begin{figure}[!htpb]
    \centering
    \includegraphics[height=65mm,width=1\columnwidth]{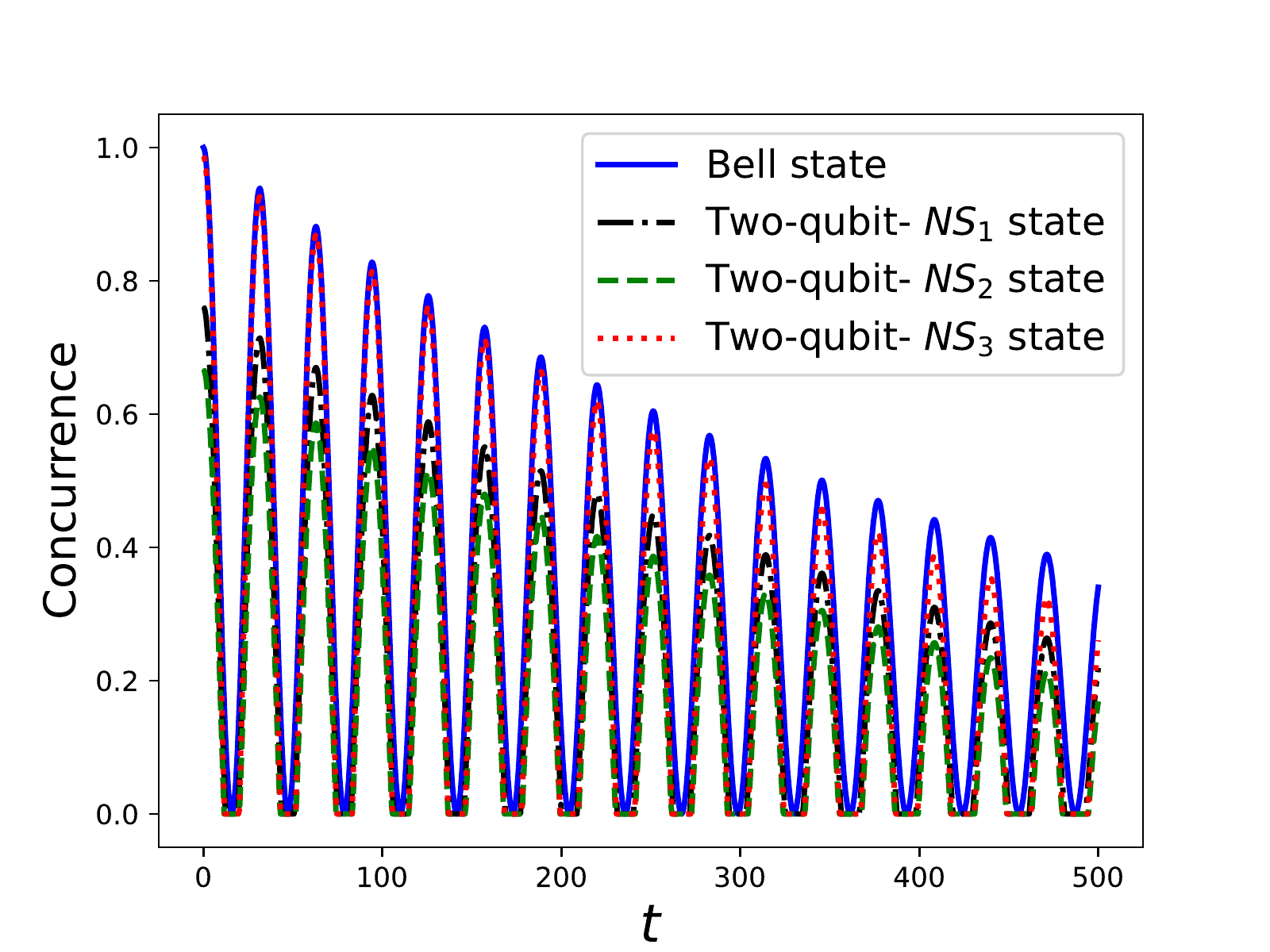}
    \caption{Concurrence variation for the two-qubit's $NS_1$, $NS_2$, $NS_3$ states, and Bell state under non-Markovian RTN noise with time. For $\gamma = 0.001$ and $b = 0.05$.}
    \label{concurNMRTN}
\end{figure}
\begin{figure}[!htpb]
    \centering
    \includegraphics[height=65mm,width=1\columnwidth]{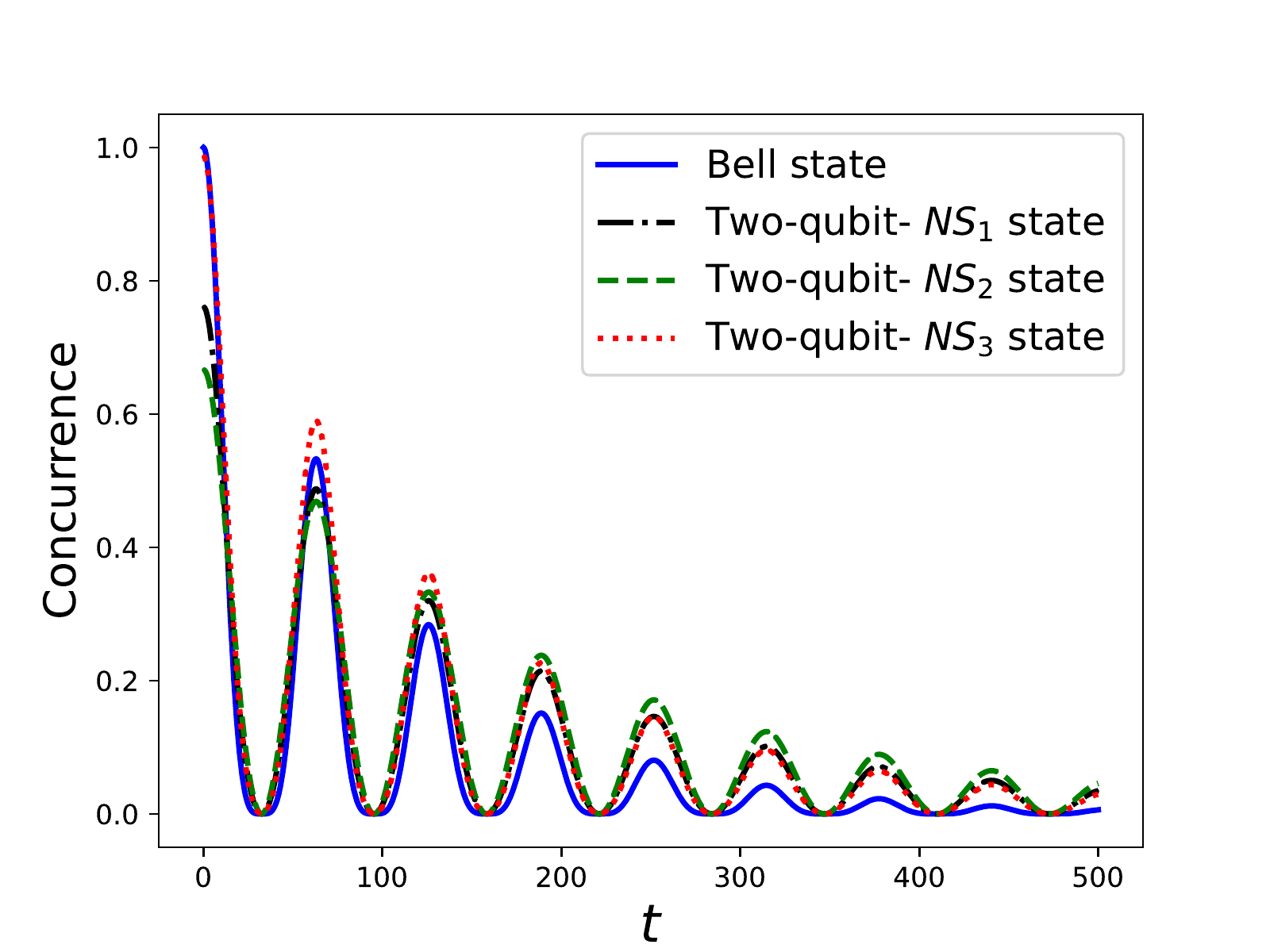}
    \caption{Concurrence variation for the two-qubit's $NS_1$, $NS_2$, $NS_3$ states, and Bell state under non-Markovian AD noise with time. For $\gamma = 1$, $g = 0.005$.}
    \label{concurNMAD}
\end{figure}

\section{\label{fid}Teleportation fidelity variation using DWFs under different noisy channels}
Quantum teleportation uses two-qubit entangled states as a resource, and the teleportation fidelity \cite{horodecki1996teleportation} is determined as
\begin{equation}
   \begin{aligned}
     F(\pmb{\rho}_{AB}) = \frac{1}{2}\left(1 + \frac{N_F(\pmb{\rho}_{AB})}{3}\right),
      \end{aligned}\label{teleportationfidelity}
\end{equation}
where $\pmb{\rho}_{AB}$ is as in Sec. \ref{two-qubit}, and $N_{F}(\pmb{\rho}_{AB}) = Tr\sqrt{T^{\dagger}T}$. The two-qubit state is advantageous for quantum teleportation iff $N_{F}(\pmb{\rho}_{AB}) > 1$, that is, $F(\pmb{\rho}_{AB}) > \frac{2}{3}$ (classical limit). Using the DWF form of states, Eq. (\ref{rho-decomposition-in-A}), we estimate the correlation matrix $T$ for the two-qubit's $NS_1$, $NS_2$, and $NS_3$ states, and Bell states. The correlation matrix elements for the two qubit's $NS_{1}$ state are 

\begin{align}
t_{11} =& 1 - 2\left(W_{1,1} + W_{1,2} + W_{1,3} + W_{1,4}\right.  \nonumber \\ &+\left. W_{4,1} + W_{4,2} + W_{4,3} + W_{4,4}\right),\nonumber\\
t_{12} =& 1 - 2\left(W_{1,2} + W_{1,4} + W_{2,1} + W_{2,3}\right.  \nonumber \\ &+\left. W_{3,1} + W_{3,3} + W_{4,2} + W_{4,4}\right),\nonumber\\
t_{13} =& 1 - 2\left(W_{1,2} + W_{1,4} + W_{2,2} + W_{2,4}\right.  \nonumber \\ &+\left. W_{3,1} + W_{3,3} + W_{4,1} + W_{4,3}\right),\nonumber\\
t_{21} =& 1 - 2\left(W_{1,1} + W_{1,2} + W_{2,3} + W_{2,4}\right.  \nonumber \\ &+\left. W_{3,3} + W_{3,4} + W_{4,1} + W_{4,2}\right),\nonumber\\
t_{22} =& 1 - 2\left(W_{1,2} + W_{1,3} + W_{2,1} + W_{2,4}\right.  \nonumber \\ &+\left. W_{3,1} + W_{3,4} + W_{4,2} + W_{4,3}\right),\nonumber\\
t_{23} =& 1 - 2\left(W_{1,2} + W_{1,3} + W_{2,2} + W_{2,3}\right.  \nonumber \\ &+\left. W_{3,1} + W_{3,4} + W_{4,1} + W_{4,4}\right),\nonumber\\
t_{31} =& 1 - 2\left(W_{1,1} + W_{1,2} + W_{2,3} + W_{2,4}\right.  \nonumber \\ &+\left.  W_{3,1} + W_{3,2} + W_{4,3} + W_{4,4}\right),\nonumber\\
t_{32} =& 1 - 2\left(W_{1,2} + W_{1,3} + W_{2,1} + W_{2,4}\right.  \nonumber \\ &+\left.  W_{3,2} + W_{3,3} + W_{4,1} + W_{4,4}\right),\nonumber\\
t_{33} =& 1 - 2\left(W_{1,1} + W_{1,4} + W_{2,1} + W_{2,4}\right.  \nonumber \\ &+\left.  W_{3,1} + W_{3,4} + W_{4,1} + W_{4,4}\right).\nonumber\\
\label{T-matrix}
\end{align}

Teleportation fidelity is calculated using Eq. (\ref{teleportationfidelity}) and the above Eq. (\ref{T-matrix}) correlation matrix elements to find the $N_{F}(\pmb{\rho}_{AB})$. Figures (\ref{fidelityNMRTN}) and (\ref{fidelityNMAD}) depict the variation of fidelity under non-Markovian RTN and non-Markovian AD noisy channels, respectively. The fidelity of the $NS_1$, $NS_2$, and $NS_3$ state and the Bell state show decaying oscillations in synchronization, with the only difference that the $NS_1$, $NS_2$, and $NS_3$ states are going below the upper bound of classical teleportation under the action of non-Markovian RTN channel as shown in Fig. (\ref{fidelityNMRTN}). From Fig. (\ref{fidelityNMAD}), it can be seen that under non-Markovian AD noise, the fidelity of the $NS_3$ state is initially similar to the fidelity of the Bell state. In contrast, $NS_1$ and $NS_2$ states have lesser values. But at longer times, the fidelity of the $NS_1$ and $NS_2$ states is similar to the fidelity of the Bell state. In contrast, the $NS_3$ state has lesser teleportation fidelity.
\begin{figure}[!htpb]
    \centering
    \includegraphics[height=65mm,width=1\columnwidth]{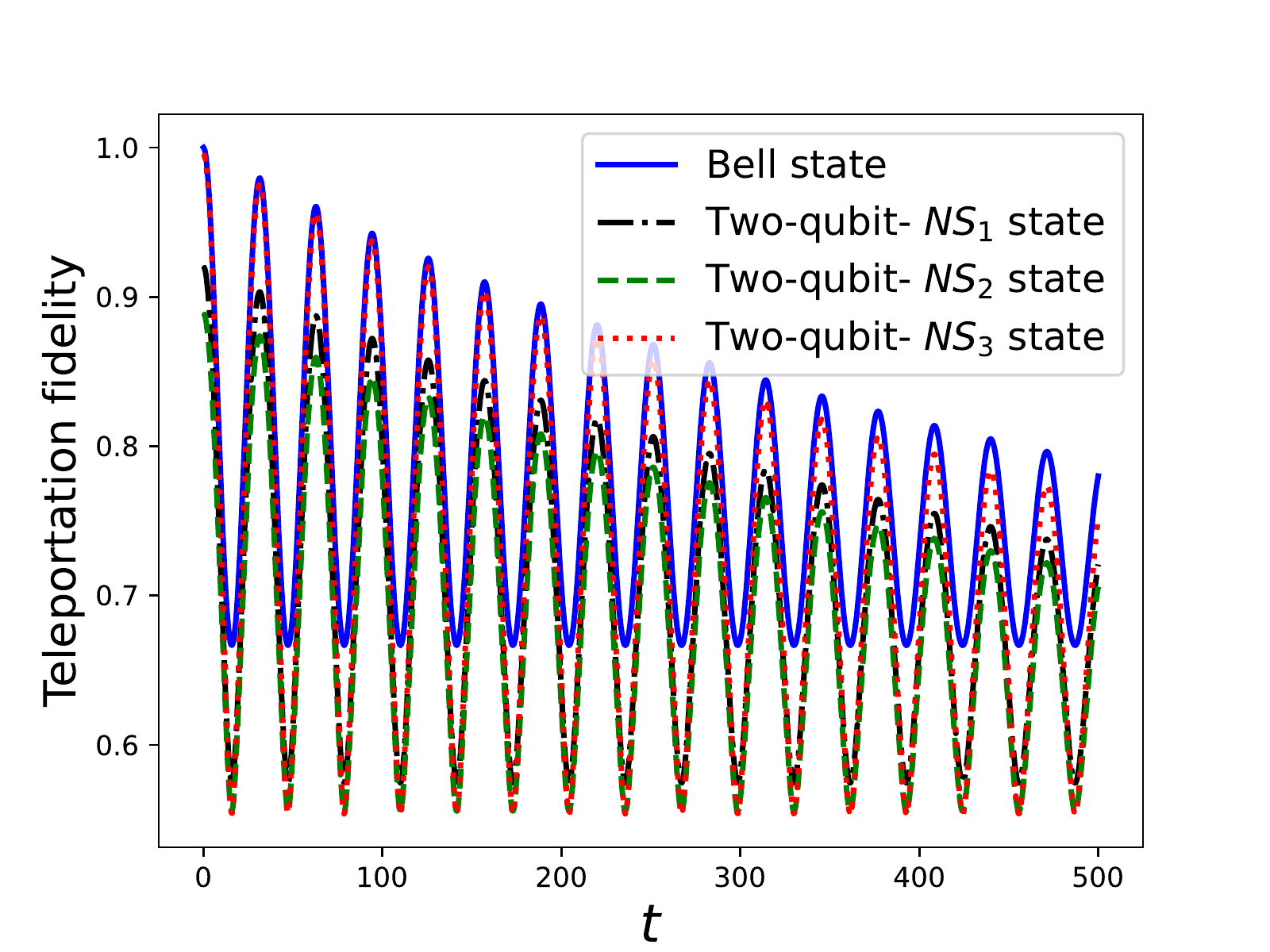}
    \caption{Variation of teleportation fidelity for the two-qubit's $NS_1$, $NS_2$, $NS_3$ states, and Bell state under non-Markovian RTN noise with time. For $\gamma = 0.001$ and $b = 0.05$.}
    \label{fidelityNMRTN}
\end{figure}
\begin{figure}[!htpb]
    \centering
    
    \includegraphics[height=65mm,width=1\columnwidth]{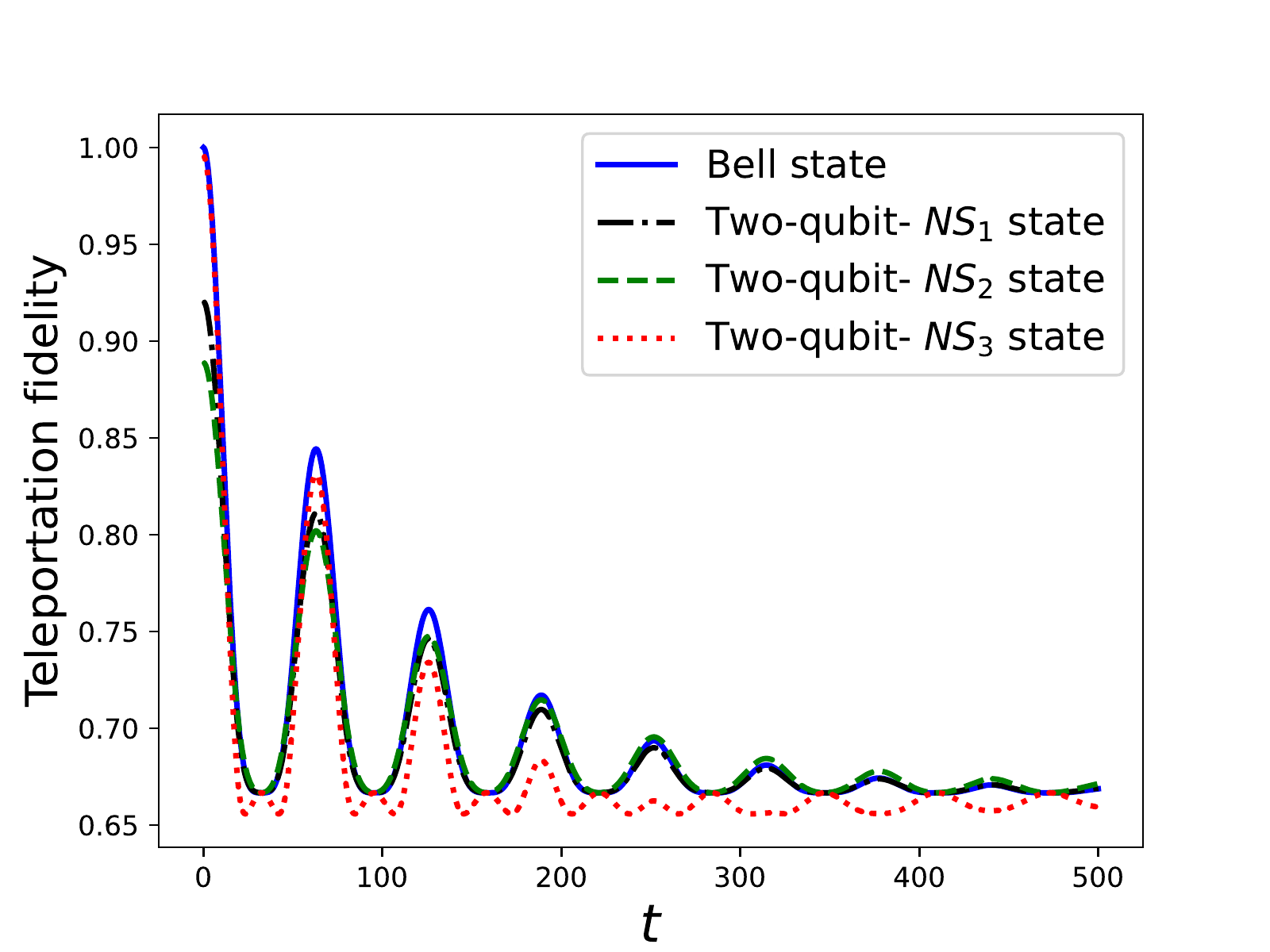}
    \caption{Variation of teleportation fidelity for the two-qubit's $NS_1$, $NS_2$, $NS_3$ states, and Bell state under non-Markovian AD noise with time. For $\gamma = 1 , g = 0.005$.}
    \label{fidelityNMAD}
\end{figure}
\section{\label{conclusion}Conclusion} 
The use of the discrete Wigner function  to investigate the behavior of quantum states under different noisy channels is significant and could provide valuable insights into the robustness of quantum information in noisy environments. The behaviour of DWFs of a qubit, qutrit, and two-qubit maximally negative quantum states was studied under different (non)-Markovian channels. Also studied was the variation of mana for qutrit's $NS_1$ and $NS_2$ states under (non)-Markovian evolution. Initially, the mana value of the qutrit's $NS_1$ state is higher than that of the $NS_2$. Yet, compared to the qutrit's $NS_2$ state, $NS_1$ state quickly dissipates under non-Markovian AD. As a result, the qutrit's $NS_2$ state endures for a longer time. Both negative quantum states of qutrit exhibit the anticipated oscillatory behaviour under the non-Markovian RTN, which is persistent for a far more extended period than the non-Markovian AD. An interesting facet of this work was the behaviour of the negative quantum states as compared to the Bell states for important quantum information aspects such as quantum coherence, entanglement, and teleportation fidelity. We investigated the quantum coherence, entanglement, and teleportation fidelity variation of the $NS_1$, $NS_2$, and $ NS_3$ states of the two-qubit system and compared this to the Bell state under different noisy channels using DWFs. Under the non-Markovian AD noise, the quantum coherence and entanglement of two-qubit $NS_1$, $NS_2$, and $NS_3$ states sustain for longer than the Bell state, but for teleportation fidelity, they behave like the Bell state. Moreover, the quantum coherence of the two-qubit $NS_1$, $NS_2$, and $NS_3$ states persist for longer than the Bell state, but for the entanglement and teleportation fidelity, the Bell states dominate under noisy dephasing channels such as depolarising and non-Markovian RTN.
\section*{Acknowledgements}
SB acknowledges the support from the Interdisciplinary Cyber-Physical Systems (ICPS) programme of
the Department of Science and Technology (DST), India,
Grant No.: DST/ICPS/QuST/Theme-1/2019/6. SB
also acknowledges support from the Interdisciplinary Research Platform (IDRP) on Quantum Information and
Computation (QIC) at IIT Jodhpur.
\appendix
\section{\label{appen-2-qubit}DWFs of two-qubit systems}
DWFs of two-qubit systems using Eq. (\ref{DWFformula}) for a particular association of MUBs given in the TABLE \ref{table3}  and (\ref{2-qubit-rho}) are given below.
\begin{align}
W_{1,1} &= \frac{1}{16} \left( 1 - a_1 - a_2 + a_3 - s_1 + s_2 + s_3 + t_{11} - t_{12}\right.\nonumber\\
&- \left. t_{13} + t_{21} - t_{22} - t_{23} - t_{31} + t_{32} + t_{33}\right),\nonumber\\
W_{1,2} &= \frac{1}{16} \left( 1 - a_1 - a_2 + a_3 - s_1 - s_2 - s_3 + t_{11} + t_{12}\right.\nonumber\\
&+ \left. t_{13} + t_{21} + t_{22} + t_{23} - t_{31} - t_{32} - t_{33}\right),\nonumber
\end{align}
\begin{align}
W_{1,3} &= \frac{1}{16} \left( 1 - a_1 + a_2 - a_3 - s_1 + s_2 + s_3 + t_{11} - t_{12}\right.\nonumber\\
&- \left. t_{13} - t_{21} + t_{22} + t_{23} + t_{31} - t_{32} - t_{33}\right),\nonumber\\
W_{1,4} &= \frac{1}{16} \left( 1 - a_1 + a_2 - a_3 - s_1 - s_2 - s_3 + t_{11} + t_{12}\right.\nonumber\\
&+ \left. t_{13} - t_{21} - t_{22} - t_{23} + t_{31} + t_{32} + t_{33}\right),\nonumber
\end{align}
\begin{eqnarray}
W_{2,1} &=& \frac{1}{16} \left( 1 - a_1 - a_2 + a_3 + s_1 - s_2 + s_3 - t_{11} + t_{12}\right.\nonumber\\
&-& \left. t_{13} - t_{21} + t_{22} - t_{23} + t_{31} - t_{32} + t_{33}\right),\nonumber\\
W_{2,2} &=& \frac{1}{16} \left( 1 - a_1 - a_2 + a_3 + s_1 + s_2 - s_3 - t_{11} - t_{12}\right.\nonumber\\
&+& \left. t_{13} - t_{21} - t_{22} + t_{23} + t_{31} + t_{32} - t_{33}\right),\nonumber\\
W_{2,3} &=& \frac{1}{16} \left( 1 - a_1 + a_2 - a_3 + s_1 - s_2 + s_3 - t_{11} + t_{12}\right.\nonumber\\
&-& \left. t_{13} + t_{21} - t_{22} + t_{23} - t_{31} + t_{32} - t_{33}\right),\nonumber\\
W_{2,4} &=& \frac{1}{16} \left( 1 - a_1 + a_2 - a_3 + s_1 + s_2 - s_3 - t_{11} - t_{12}\right.\nonumber\\
&+& \left. t_{13} + t_{21} + t_{22} - t_{23} - t_{31} - t_{32} + t_{33}\right),\nonumber
\end{eqnarray}
\begin{eqnarray}
W_{3,1} &=& \frac{1}{16} \left( 1 + a_1 + a_2 + a_3 - s_1 + s_2 + s_3 - t_{11} + t_{12}\right.\nonumber\\
&+& \left. t_{13} - t_{21} + t_{22} + t_{23} - t_{31} + t_{32} + t_{33}\right),\nonumber\\
W_{3,2} &=& \frac{1}{16} \left( 1 + a_1 + a_2 + a_3 - s_1 - s_2 - s_3 - t_{11} - t_{12}\right.\nonumber\\
&-& \left. t_{13} - t_{21} - t_{22} - t_{23} - t_{31} - t_{32} - t_{33}\right),\nonumber\\
W_{3,3} &=& \frac{1}{16} \left( 1 + a_1 - a_2 - a_3 - s_1 + s_2 + s_3 - t_{11} + t_{12}\right.\nonumber\\
&+& \left. t_{13} + t_{21} - t_{22} - t_{23} + t_{31} - t_{32} - t_{33}\right),\nonumber\\
W_{3,4} &=& \frac{1}{16} \left( 1 + a_1 - a_2 - a_3 - s_1 - s_2 - s_3 - t_{11} - t_{12}\right.\nonumber\\
&-& \left. t_{13} + t_{21} + t_{22} + t_{23} + t_{31} + t_{32} + t_{33}\right),\nonumber
\end{eqnarray}
\begin{eqnarray}
W_{4,1} &=& \frac{1}{16} \left( 1 + a_1 + a_2 + a_3 + s_1 - s_2 + s_3 + t_{11} - t_{12}\right.\nonumber\\
&+& \left. t_{13} + t_{21} - t_{22} + t_{23} + t_{31} - t_{32} + t_{33}\right),\nonumber\\
W_{4,2} &=& \frac{1}{16} \left( 1 + a_1 + a_2 + a_3 + s_1 + s_2 - s_3 + t_{11} + t_{12}\right.\nonumber\\
&-& \left. t_{13} + t_{21} + t_{22} - t_{23} + t_{31} + t_{32} - t_{33},\right)\nonumber\\
W_{4,3} &=& \frac{1}{16} \left( 1 + a_1 - a_2 - a_3 + s_1 - s_2 + s_3 + t_{11} - t_{12}\right.\nonumber\\
&+& \left. t_{13} - t_{21} + t_{22} - t_{23} - t_{31} + t_{32} - t_{33}\right),\nonumber\\
W_{4,4} &=& \frac{1}{16} \left( 1 + a_1 - a_2 - a_3 + s_1 + s_2 - s_3 + t_{11} + t_{12}\right.\nonumber\\
&-& \left. t_{13} - t_{21} - t_{22} + t_{23} - t_{31} - t_{32} + t_{33}\right).\nonumber\\
\end{eqnarray}

\bibliography{apssamp}
\end{document}